\documentclass{pasj02}
\usepackage[switch,mathlines]{lineno}
\Received{$\langle$reception date$\rangle$}
\Accepted{$\langle$acception date$\rangle$}
\Published{$\langle$publication date$\rangle$}
\usepackage{float}
\usepackage{comment}
\usepackage{subcaption}
\usepackage{graphicx}
\usepackage{afterpage}
\usepackage{forest}
\usepackage{multirow}
\usepackage{booktabs}
\usepackage{placeins}
\usepackage{threeparttable}
\usepackage{natbib}
\usepackage{xcolor}
\definecolor{xlinkcolor}{rgb}{0, 0, 1}
\usepackage[bookmarks=true, pdfnewwindow=true, colorlinks=true, linkcolor=xlinkcolor, citecolor=xlinkcolor, filecolor=xlinkcolor, urlcolor=xlinkcolor, final=true]{hyperref}
\usepackage{bookmark}

\usepackage{ulem}
\begin{document}
%\linenumbers
\title{ XRISM Reveals a Kinematically Coherent Core System \\of the Nearby Cool-Core Cluster Abell~2199}

\author{
Kazunori~\textsc{Suda},\altaffilmark{1}\orcid{0009-0003-3920-131X}
Kyoko~\textsc{Matsushita},\altaffilmark{1}\orcid{0000-0003-2907-0902}
Kosuke~\textsc{Sato},\altaffilmark{2}\orcid{0000-0001-5774-1633}
Kotaro~\textsc{Fukushima},\altaffilmark{1}\orcid{0000-0001-8055-7113}
Ming~\textsc{Sun},\altaffilmark{3}\orcid{0000-0001-5880-0703}
Caroline~\textsc{Kilbourne},\altaffilmark{4}\orcid{0000-0001-9464-4103}
John A.~\textsc{ZuHone}, \altaffilmark{5}\orcid{0000-0003-3175-2347}
Edmund~\textsc{Hodges-Kluck}, \altaffilmark{4}\orcid{0000-0002-2397-206X}
Shogo B.~\textsc{Kobayashi},\altaffilmark{6}\orcid{0000-0001-7773-9266}
Simon~\textsc{Dupourqué},\altaffilmark{7}\orcid{0000-0003-2715-8986} 
Daniel R.~\textsc{Wik},\altaffilmark{8}\orcid{0000-0001-9110-2245}
Priyanka~\textsc{Chakraborty},\altaffilmark{5,9}\orcid{0000-0002-4469-2518}
Arnab~\textsc{Sarkar},\altaffilmark{9,10}\orcid{0000-0002-5222-1337}
Marie~\textsc{Kondo},\altaffilmark{11}\orcid{0009-0005-5685-1562}
Itsuki~\textsc{Aihara},\altaffilmark{1}\orcid{0009-0009-8380-1260}
Eric D.~\textsc{Miller}, \altaffilmark{10}\orcid{0000-0002-3031-2326}
and
Fran\c{c}ois~\textsc{Mernier}\altaffilmark{7}\orcid{0000-0002-7031-4772}
}

\altaffiltext{1}{Department of Physics, Tokyo University of Science, 1-3 Kagurazaka, Shinjuku-ku, Tokyo 162-8601, Japan}
\altaffiltext{2}{Department of Astrophysics and Atmospheric Sciences, Kyoto Sangyo University, Motoyama, Kamigamo, Kita-ku, Kyoto, Kyoto 603-8555, Japan}
\altaffiltext{3}{Department of Physics \& Astronomy, University of Alabama in Huntsville, 301 Sparkman Dr NW, Huntsville, AL, 35899, USA}
\altaffiltext{4}{NASA Goddard Space Flight Center, 8800 Greenbelt Rd., Greenbelt, MD 20771, USA}
\altaffiltext{5}{Center for Astrophysics $\vert$ Harvard \& Smithsonian, 60 Garden Street, Cambridge, MA, 02138}
\altaffiltext{6}{Department of Physics, Rikkyo University, 3-34-1 Nishi Ikebukuro, Toshima-ku, Tokyo 171-8501, Japan}
\altaffiltext{7}{Univ Toulouse, CNES, CNRS, IRAP, Toulouse, France}
\altaffiltext{8}{Department of Physics \& Astronomy, University of Utah, 270 South 1400 East, Salt Lake City, UT, 84112, USA}
\altaffiltext{9}{Department of Physics, University of Arkansas, 825 W Dickson st., Fayetteville, AR 72701, USA}
\altaffiltext{10}{Kavli Institute for Astrophysics and Space Research, Massachusetts Institute of Technology, 70 Vassar St, Cambridge, MA 02139}
\altaffiltext{11}{Department of Physics, Saitama University, 255 Shimo-Okubo, Sakura-ku, Saitama, Saitama 338-8570, Japan}
\email{kazunori.suda.gc@gmail.com, matusita@rs.tus.ac.jp}

%\KeyWords{key word --- key word --- \dots --- key word }

\maketitle

\begin{abstract} 

We present the results of a deep 251 ks XRISM/Resolve observation of the cool core of the galaxy cluster Abell~2199.
From the integrated spectrum of the central $3' \times 3'$ Resolve field of view ($104 \times 104 \mathrm{~kpc}^2$), 
we find that
 the intracluster medium (ICM) redshift is consistent with that of the brightest cluster galaxy,
 within the optical-redshift uncertainty.
 This indicates that they form a kinematically coherent core system, which is 
 offset from the mean cluster redshift by
$\sim200~\mathrm{km~s^{-1}}$.
The observed velocity dispersion of $\sim100~\mathrm{km~s^{-1}}$
corresponds to a three-dimensional Mach number of $\mathcal{M}_{\mathrm{3D}}=0.16$ and a non-thermal pressure fraction of $P_{\mathrm{NT}}/P_{\mathrm{tot}}=1.4\pm0.2$\%.
Abell~2199 is one of the most dynamically quiescent relaxed clusters observed with XRISM, despite the presence of radio jets and a plume-like structure possibly associated with sloshing motions.
Order-of-magnitude estimates suggest that turbulent dissipation could offset a non-negligible fraction of the radiative cooling losses, 
with $Q_{\mathrm{turb}}/Q_{\mathrm{cool}}\approx0.1 \color{black}$ for a large-scale driver such as sloshing and larger values for smaller AGN-feedback scales. 
Finally, we detect a  localized enhancement of the Fe~\textsc{xxv} He$\alpha$ $y$ line in the southeast region, which spatially coincides with a \textit{Chandra} surface brightness discontinuity. 

\end{abstract}

\section{Introduction}
Galaxy clusters are the most massive gravitationally bound systems in the Universe and form hierarchically within the cold dark matter (CDM) framework \citep{kravtsov_formation_2012}. 
They grow through the continuous accretion of dark matter and baryons along surrounding filaments, where accretion shocks heat the intracluster medium (ICM) to X-ray-emitting temperatures \citep{kravtsov_formation_2012}.

Ongoing cluster growth can also leave observable imprints in cluster cores.
When a smaller cluster or group passes near the center of a more massive system, the resulting perturbation of the gravitational potential can trigger gas sloshing \citep{markevitch2007, zuhone2011}. 
In this process, the low-entropy core is displaced from the potential minimum and subsequently oscillates, driving bulk motions and turbulence in the ICM \citep{vazza2012, zuhone2013, zuhone2016}. 
These sloshing motions often produce cold fronts and, depending on the viewing geometry, spiral-like patterns in the X-ray surface brightness and temperature distribution.
Observational evidence for such spiral structures has been reported  \citep{churazov2000, ueda2017, douglass2018}, and
\citet{lagana2010} identified spiral features in roughly half of a sample of 15 nearby clusters ($0.01<z<0.06$).

In the core of relaxed clusters, the radiative cooling time of the ICM is often much shorter than the Hubble time.
An additional heat source is therefore required to maintain
the observed thermodynamic structure of cool cores.
Although feedback from the active galactic nuclei (AGN) in the 
 brightest cluster galaxies (BCG) is widely regarded as the primary self-regulating heating mechanism, gas motions induced by mergers and sloshing may also play an important role in redistributing energy within the core (e.g., \citealt{zhuravleva2018,dupourque2023}).
Understanding the interplay between merger-driven sloshing and AGN feedback is thus essential to describe the thermal balance and long-term evolution of cool cores.

High-resolution X-ray spectroscopy with the Hitomi/SXS \citep{kelley2016, takahashi2018} and XRISM/Resolve \citep{ishisaki2025, kelley2025, tashiro2025} microcalorimeters has enabled direct measurements of the line-of-sight (LOS) bulk velocities and velocity dispersions of the ICM in galaxy clusters \citep{hitomicollaboration2018a,xrism2319}. 
Resolve observations have provided new insight into the  ICM kinematics in the vicinity of BCGs and the possible contributions of  sloshing and AGN feedback
\citep{xrism_ophiuchus, rose2025, xrism_A2029_c, xrism_a2029_bcore, xrism_centaurus, xrism_a2029_core, aihara2026, xrism_centaurus_kondo, mccall2026_a3571, xrism_cygnus, sarkar2026_a1795, xrism_m87_aurora, xrism_pks, xrism_perseus, xrism_m87, congyao_perseus}. 
In particular, the Centaurus cluster exhibits LOS bulk velocities of $130$--$310~\mathrm{km~s^{-1}}$ relative to the BCG, providing the first spectroscopic evidence for large-scale gas motions consistent with sloshing \citep{xrism_centaurus}.
In the Perseus cluster, where previous X-ray observations revealed a prominent spiral structure in the core \citep{churazov2000,perseus_aurora_ccd,sanders2020},
Resolve detected LOS bulk-velocity variations of $\sim 100~\mathrm{km~s^{-1}}$ associated with the spiral cold fronts, in good agreement with sloshing simulations \citep{xrism_perseus}.
By contrast, in several other cool-core clusters (Ophiuchus, \citealt{xrism_ophiuchus}; Hydra A, \citealt{rose2025}; Center and N2 pointings of A2029, \citealt{xrism_a2029_bcore}; M87, \citealt{xrism_m87_aurora}; Cygnus A, \citealt{xrism_cygnus}; A3571, \citealt{aihara2026}; A1795, \citealt{sarkar2026_a1795}), the LOS velocity offsets relative to the BCG are generally smaller than $100~\mathrm{km~s^{-1}}$. 

The observed velocity dispersions in cool cores are also typically modest,  $\sim 100$--$200~\mathrm{km~s^{-1}}$.
Enhanced velocity dispersion toward the centers of Perseus and M87 suggests the presence of localized gas motions likely associated with AGN feedback.
In Cygnus A, the large velocity dispersion of $\sim 260~\mathrm{km~s^{-1}}$,  with its exceptionally powerful radio source, also indicates strong AGN-driven disturbances.
Meanwhile, comparisons with numerical simulations have suggested that the velocity dispersions measured with XRISM in cool-core clusters are systematically lower than expected \citep{xrism_cosmosim}.
The corresponding kinetic-to-total pressure ratio is also smaller than the simulated values, with a median of 2.2\%, indicating that the cool cores are dynamically quieter than predicted by current simulations.

% Chandra Map
\begin{figure*}
\includegraphics[width=0.48\linewidth]{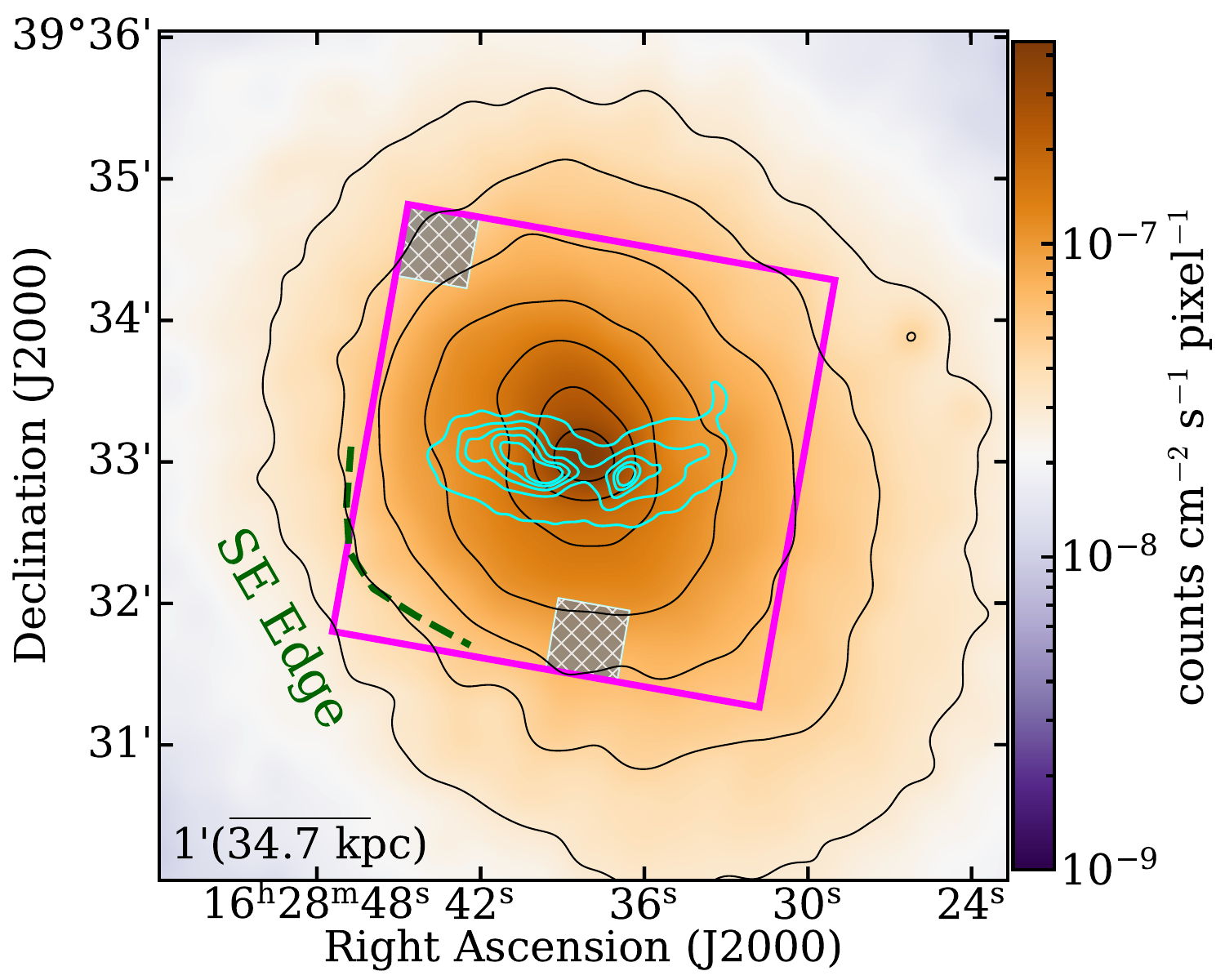}\includegraphics[width=0.48\linewidth]{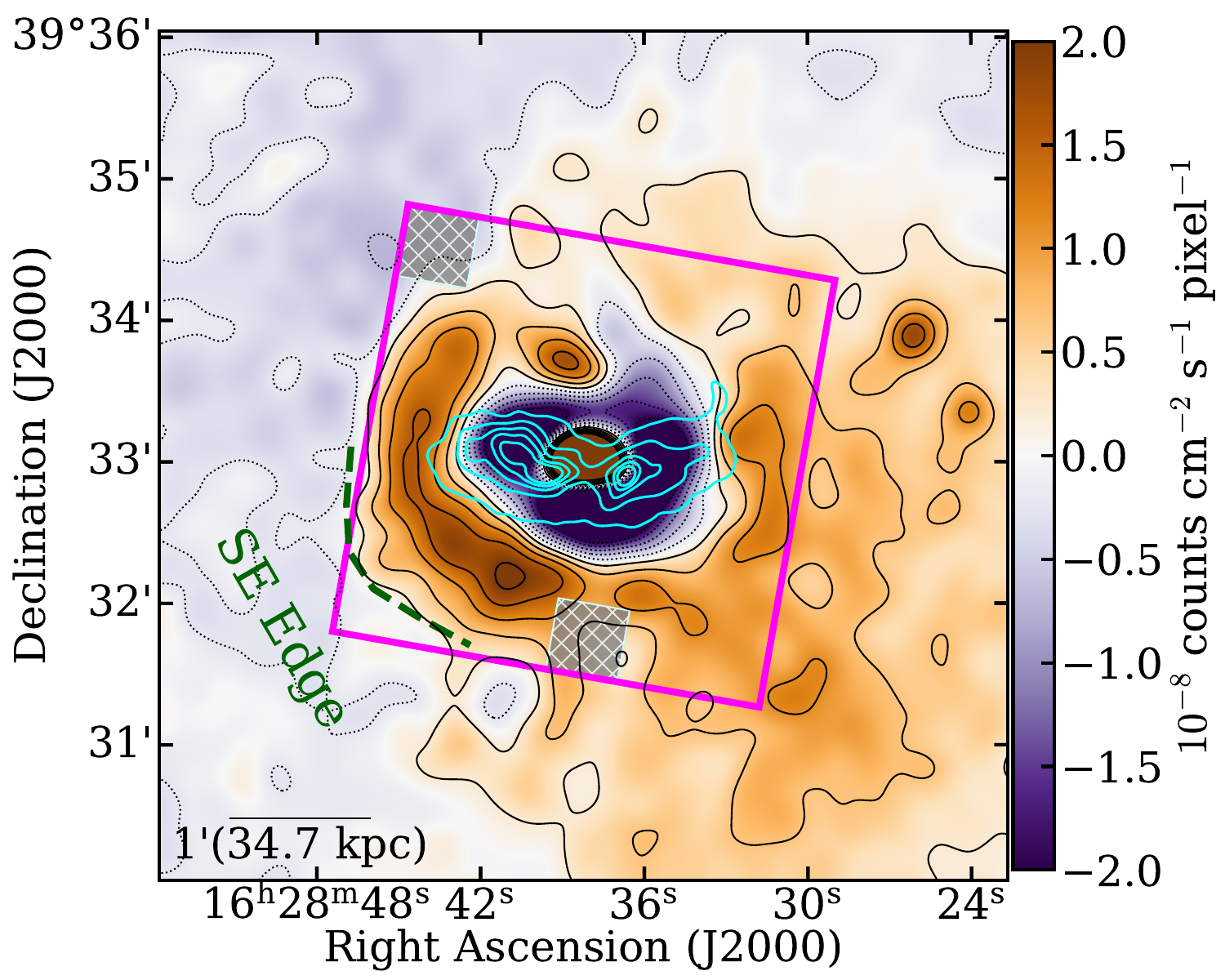}
\caption{
(left panel) \textit{Chandra} 2.0--8.0 keV image of the central region of Abell~2199. The image has been smoothed with a Gaussian kernel of $\sigma \approx 7.4^{\prime\prime}$ (15  pixels).
Contours are drawn from $3 \times 10^{-8}$ to $5 \times 10^{-7}$ $\mathrm{counts~cm^{-2}~s^{-1}~pixel^{-1}}$ in 8 logarithmically spaced levels.  The square indicates the Resolve FOV, with the two hatched boxes corresponding to the pixels excluded from our analysis.
The dashed curve shows the position of a surface-brightness edge \citep{nulsen2013}. Cyan contours represent the 120-168 MHz radio intensity obtained from the LOFAR Two-metre Sky Survey Third Data Release (LoTSS DR3; \citealt{shimwell2026}, Mosaic number=588). The radio contours are drawn from $0.1$ to $1.0$ $\mathrm{Jy~beam}^{-1}$ in $5$ linearly spaced levels.
(right panel) Residual image obtained by subtracting the best-fit elliptical $\beta$ model from the data shown in the left panel. Contours are drawn from $-2\times 10^{-8}$ to $2\times 10^{-8}$ $\mathrm{counts~cm^{-2}~s^{-1}~pixel^{-1}}$ in 10 linearly spaced levels.
{Alt text: The X-ray flux and residual images centered on the brightest cluster galaxy of Abell 2199 are shown using the gnomonic projection in the J2000.0 equatorial coordinates.
The intensity distribution within each image is indicated by color and corresponding contour lines.
North is up, and east is to the left.
Right ascension is given in sexagesimal format on the abscissa, declination in decimal format on the ordinate.}
}
\label{fig:fov-chandra}
\end{figure*}

In this context, an intriguing target is the cool-core cluster Abell~2199 at $z=0.03$ \citep{johnstone2002}.  
Previous observations with Chandra and Suzaku have revealed an asymmetric X-ray surface-brightness distribution within the central $\sim2.5$ arcmin, while the outer regions appear relatively smooth and symmetric \citep{johnstone2002, sanders2006, kawaharada2010, nulsen2013}.
Although no clear spiral structure like that seen in the Perseus cluster is observed, the asymmetric surface-brightness and temperature distributions, including a plume-like feature toward the southwest and a surface-brightness edge at 1.5 arcmin to the southeast, suggest ongoing sloshing activity.

Numerical simulations replicating the known characteristic features indicate that the sloshing orbital plane is viewed nearly edge-on, as well as predict that approximately 0.8 Gyr has passed since the closest approach of the perturber \citep{machado2022}.
The dynamical state of the core is also reflected in the activity of its central galaxy.
The BCG, NGC~6166, hosts the radio source 3C~338, whose lobes extend to the east and west \citep{burns1983}. Depressions in X-ray surface brightness that coincide with radio lobes indicate interactions between radio outbursts and the surrounding ICM \citep{owen1998}. A radio ridge located $\sim10''$ south of the core may represent an older detached jet, potentially influenced by sloshing motions \citep{burns1983, nulsen2013}.

In this paper, we present the results of an XRISM observation of Abell~2199, mainly focusing on its ICM velocity structure; a dedicated study on the chemical abundances of this cluster will be presented in a forthcoming paper (Mallet et al., in prep.). 
Throughout this paper, we use $H_0$ = 70 km $\mathrm{s^{-1}}$ $\mathrm{Mpc^{-1}}$, $\Omega_m$ = 0.3 flat cosmology, in which $1' = 34.7$ kpc at the cluster redshift. All errors are in 1$\sigma$ confidence interval, otherwise noted.

\section{Observation and Data Reduction}

XRISM observed the central region of Abell~2199 from 2024 October 14 to 19 (ObsID: 2010089010), with the pointing position at (RA, Dec) = (247.15924$^\circ$, 39.55073$^\circ$; J2000).

To examine the X-ray morphology within and around the Resolve field of view (FOV), 
Figure \ref{fig:fov-chandra} shows an archival 2.0--8.0 keV Chandra/ACIS image 
(ObsID:10748; the sequence is included in the list at \url{https://doi.org/10.25574/cdc.522}) and the corresponding residual image  after subtracting  the best-fit two-dimensional elliptical $\beta$-model. 
To construct this model, 
we performed a least-squares fit to the $10' \times 10'$ Chandra image smoothed using a Gaussian kernel of $\sigma = 7.4''$.
The resulting best-fit parameters are the major and minor core radii of $a=17.1$ kpc and $b=13.8$ kpc (corresponding to an ellipticity of $\epsilon \simeq 0.20$), $\beta=0.40$\color{black}, and a major-axis position angle of $31.8^\circ$ (from north to east).
The 3.05$'\times 3.05'$ square Resolve FOV is overlaid on both images and covers roughly half of the cool core.

The X-ray peak, which nearly coincides with the BCG, 
is located near the center of the Resolve FOV.
As shown in Figure \ref{fig:fov-chandra},
a plume-like feature is visible toward the southwest, possibly corresponding to the inner portion of the sloshing spiral \citep{machado2022}.
The FOV includes the two radio lobes that extend 40$"$ to the east and west.

Resolve was operated with the gate valve closed.
The Resolve data were processed using HEASoft version 6.34 and the calibration database (CALDB) ver20240815, following standard screening criteria (\url{https://heasarc.gsfc.nasa.gov/docs/xrism/analysis/quickstart/xrism_quick_start_guide_v2p3_240918a.pdf}). 
No significant variability was detected in the light curve; therefore, no additional filtering was applied. 
The resultant cleaned exposure is 251 ks. 
For the spectral analysis, we selected only high-resolution events (\texttt{TYPE=Hp}) from the full Resolve FOV, excluding pixel 12 (calibration pixel) and pixel 27, which is affected by a known gain anomaly.
We generated the redistribution matrix file (RMF) using the \texttt{rslmkrmf} task 
with the \texttt{whichrmf} parameter set to \texttt{L},
and the ancillary response files (ARFs) using the \texttt{xaarfgen} task 
  with \texttt{sourcetype} set to \texttt{image}. 
  We confirmed that the velocity measurements presented in this paper are consistent with those obtained using the \texttt{XL} RMF. 
We used the 2--8 keV Chandra/ACIS X-ray image described above as the input surface brightness distribution for creating ARFs. This image provides a spatial template extending at least $7^\prime$ from the Resolve pointing center, which is sufficiently larger than the Resolve FOV. 

\section{Analysis and results}
\subsection{Spectral Modeling}
We performed spectral fitting using XSPEC version 12.14.1.
Model fits were evaluated by minimizing the C-statistic \citep{cash1979a}. 
The ICM emission was modeled with \texttt{bapec} (ATOMDB version 3.1.3), which describes an optically thin thermal plasma in collisional ionization equilibrium. 
In this model, the thermal broadening is determined by the electron temperature while the non-thermal one is parameterized as a LOS velocity dispersion. 
We adopted the proto-solar abundances table of \citet{lodders2009}, using \texttt{lpgs} in XSPEC. 
Foreground Galactic absorption was modeled with \texttt{tbabs}, with the neutral hydrogen column density fixed at $8.05\times 10^{19} \,\mathrm{cm}^{-2}$ \citep{bekhti2016}.

The Non-X-ray Background (NXB) was treated by including a background model component directly in the spectral fit. 
This NXB model component, consisting of a power-law continuum and Gaussian instrumental lines, follows the procedure provided by the XRISM team\footnote{\url{https://heasarc.gsfc.nasa.gov/docs/xrism/analysis/nxb/nxb_spectral_models.html}}. 

\begin{figure*}
    \centering
    \includegraphics[width=\linewidth]{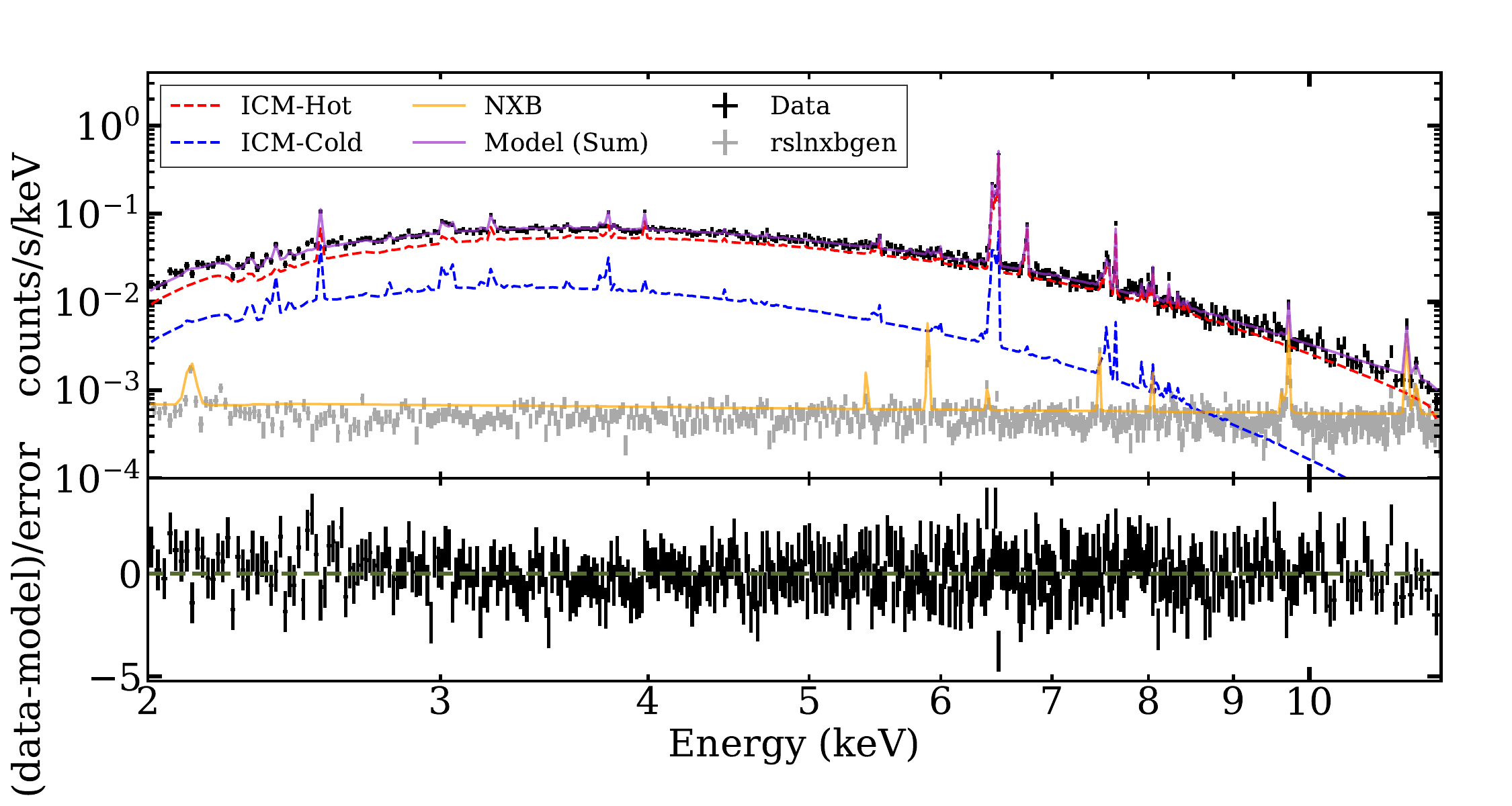}
     
    \includegraphics[width=0.48\linewidth]{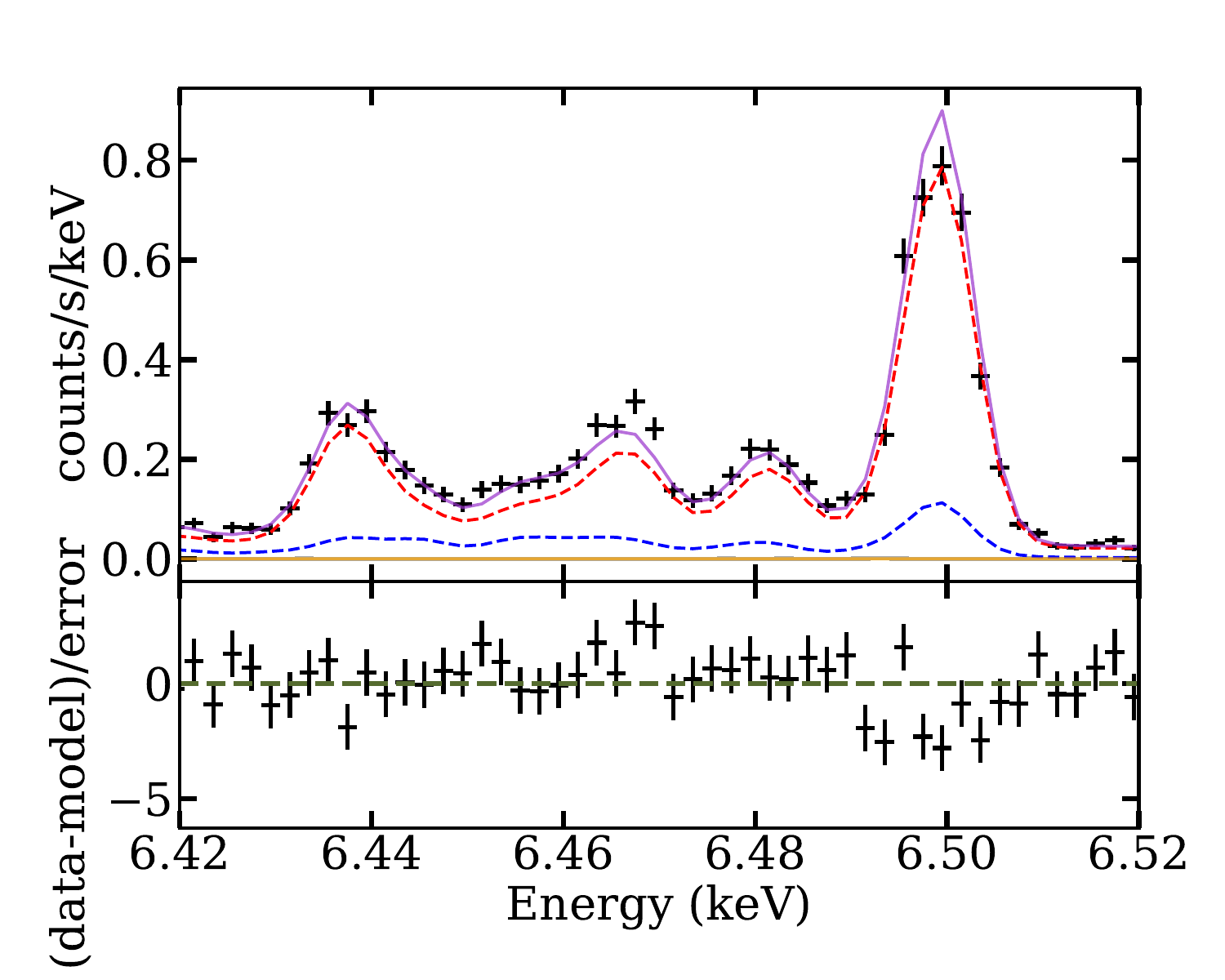}\includegraphics[width=0.48\linewidth]{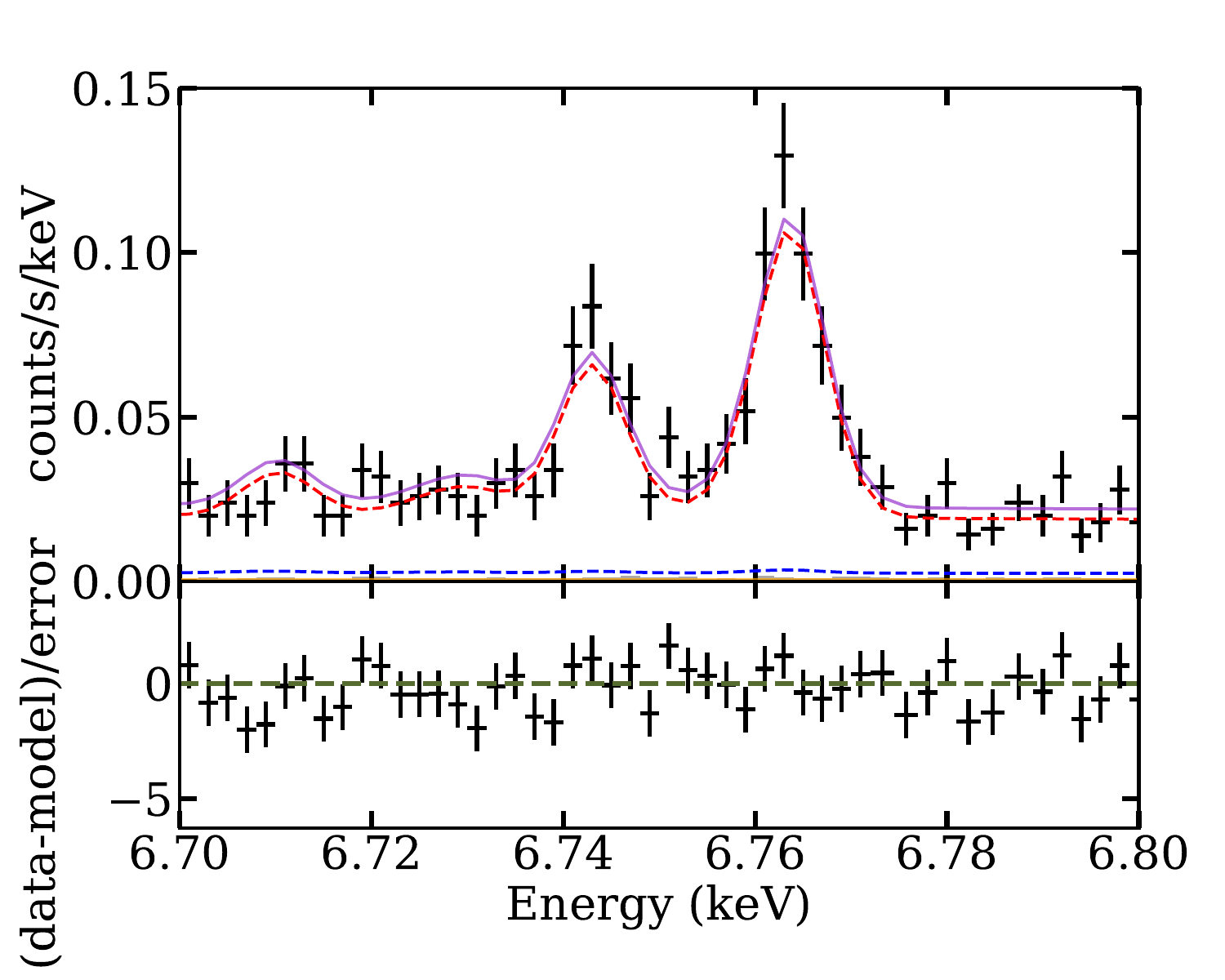}
     \caption{The best-fit FOV spectrum obtained with Resolve and the residuals for the baseline 2T model (top) 
     and enlarged views around the Fe He$\alpha$ triplet (bottom left) and Fe Ly$\alpha$ (bottom right).
     Contributions from the two thermal components (dashed lines) and the best-fit NXB model (orange solid line)  are also shown. 
     The COR-weighted NXB spectrum generated using the \texttt{rslnxbgen} tool is also shown in gray for comparison. 
     Alt text: The Resolve spectrum and best-fit two-temperature model for Resolve's field of view are shown in three panels by data points and model lines. The top one is given in the two to twelve kiloelectronvolt band.
     The two bottom ones are in the six-point-four-two- to six-point-five-two-kiloelectronvolt and six-point-seven- to six-point-eight-kiloelectronvolt bands, focusing on the iron-helium alpha and Lyman alpha complexes, respectively. In each panel, the upper panel shows the data points in units of count per second per kiloelectronvolt, and the residuals are shown in the lower panel, both of which are plotted against photon energy in units of kiloelectronvolt. 
     Contributions from thermal components and the background model are indicated, and the simulated non-X-ray background spectrum is also overlaid to demonstrate consistency with the background model.
     }

    \label{fig:fov-spectrum}
\end{figure*}

We alternatively generated an NXB spectrum weighted by the cut-off rigidity (COR)
 using the \texttt{rslnxbgen} tool to evaluate the systematic effects of the NXB treatment.
As shown in section 3.3.1 (Figure \ref{fig:fov-spectrum}), the normalization of our best-fit NXB model is approximately 20\% higher than that of the \texttt{rslnxbgen} spectrum. 
 Nevertheless, 
both approaches of including the NXB model and replicating the NXB spectrum yielded consistent results within the bounds of statistical uncertainty. We adopted the former approach as our baseline analysis.

The cosmic X-ray background (CXB) was not included in our final model, as its expected count rate in the 2--12 keV energy band is estimated to be only $\sim 10$\% of the NXB count rate.  
We confirmed that including a CXB component did not significantly affect the best-fit parameters or improve the C-statistic; therefore, we report the results obtained without the CXB. 

The barycentric-corrected recessional velocity of the BCG, NGC~6166, is adopted as 9295~$\mathrm{km~s^{-1}}$ ($z_\textup{BCG}=0.031$). This systemic velocity is determined from the average stellar velocity within a 1 kpc radius of the nucleus, based on the fit to the stellar absorption lines from the SDSS/MaNGA data \citep{bundy2015}. Another velocity estimation from the emission lines, also from the MaNGA data cube, is $\sim$50~$\mathrm{km~s^{-1}}$ higher, which is consistent with the SDSS/BOSS \citep{dauson2013}  redshift of the galaxy based on emission lines ($z=0.03119$).
The published optical measurements span $9276$–$9348~\mathrm{km s^{-1}}$ \citep{Zabludoff1990, Zabludoff1993, Coziol2009, Lauer2014, Bender2015, Kluge2020}, and the adopted value is close to the median of these estimates.
The bulk velocities of the ICM in this work were measured relative to this value using $v_\textup{bulk} = c \times (z_\textup{obs} - z_\textup{BCG})/(1 + z_\textup{BCG})$.
A barycentric correction of $-12.5~\mathrm{km~s^{-1}}$ was also applied to the measured bulk velocities.

\begin{figure*}
  \centering
  \includegraphics[width=0.48\linewidth]{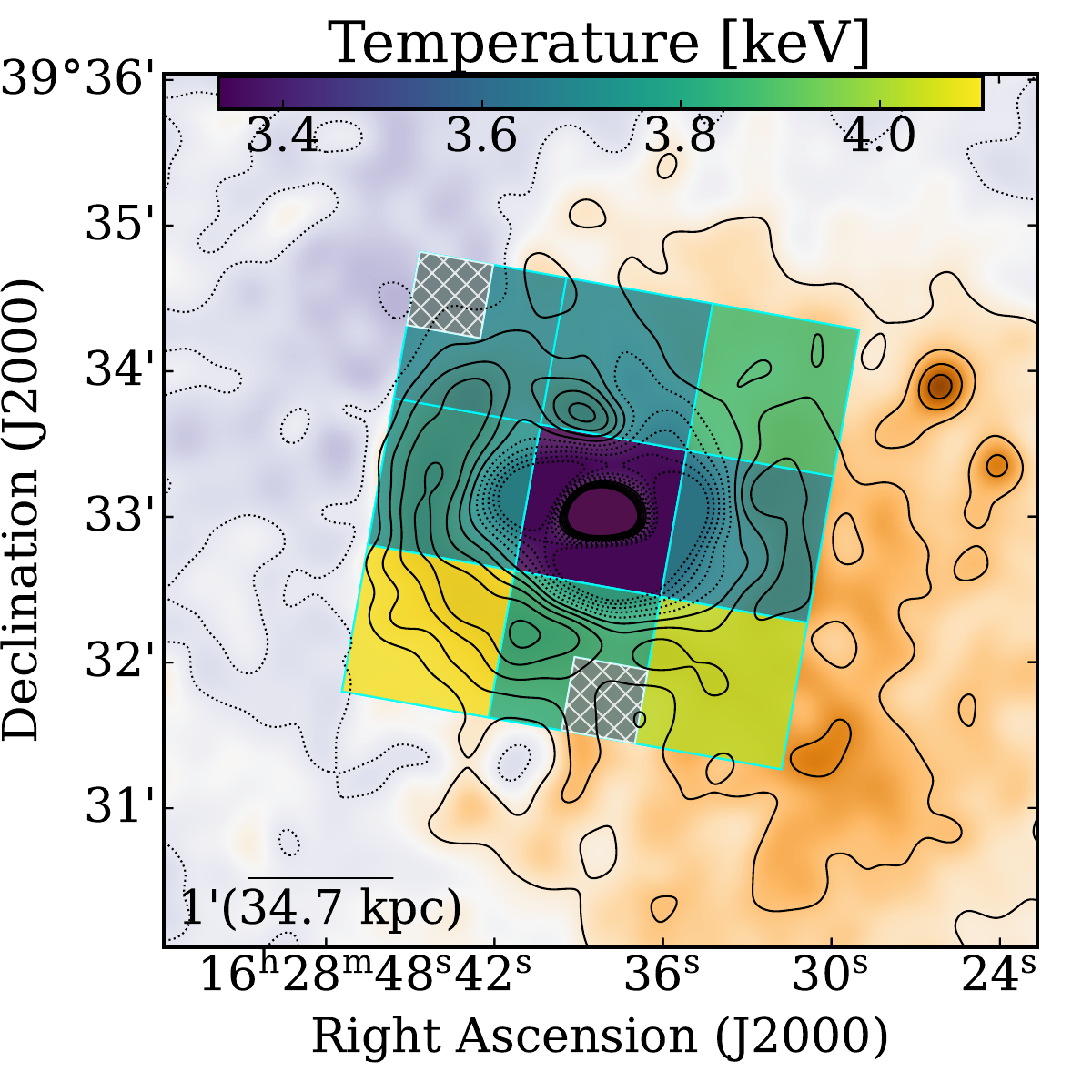}\includegraphics[width=0.48\linewidth]{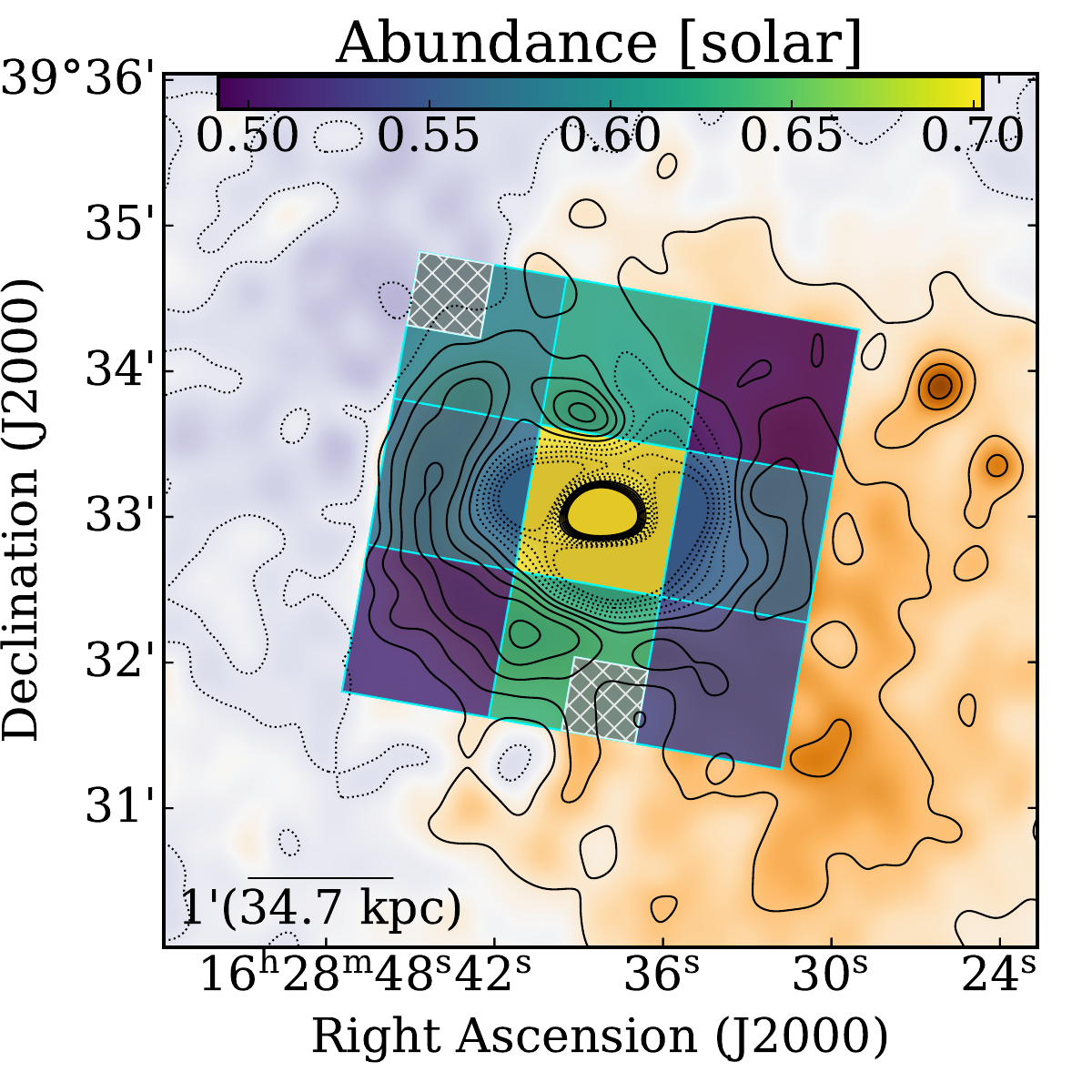}
  
  \includegraphics[width=0.48\linewidth]{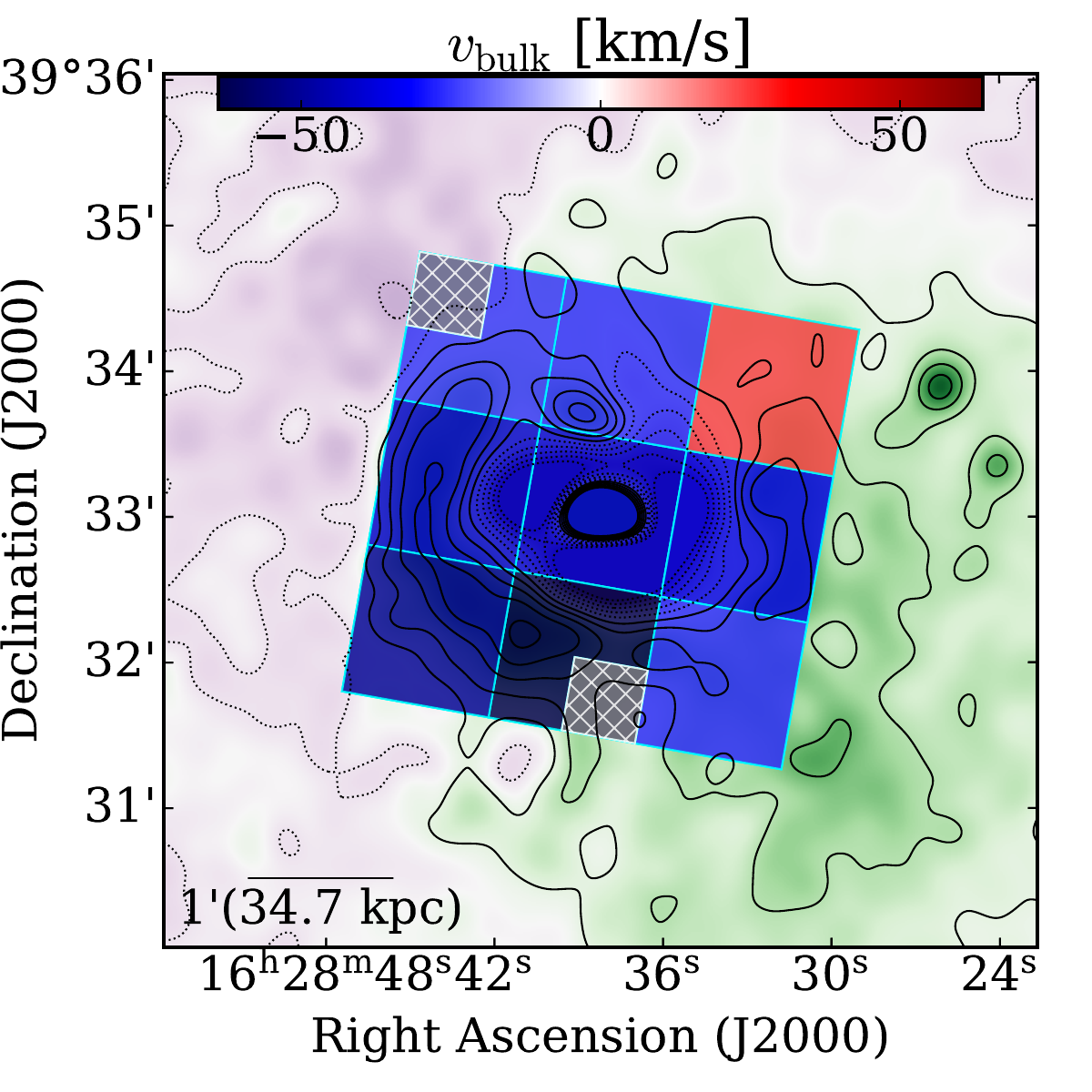}\includegraphics[width=0.48\linewidth]{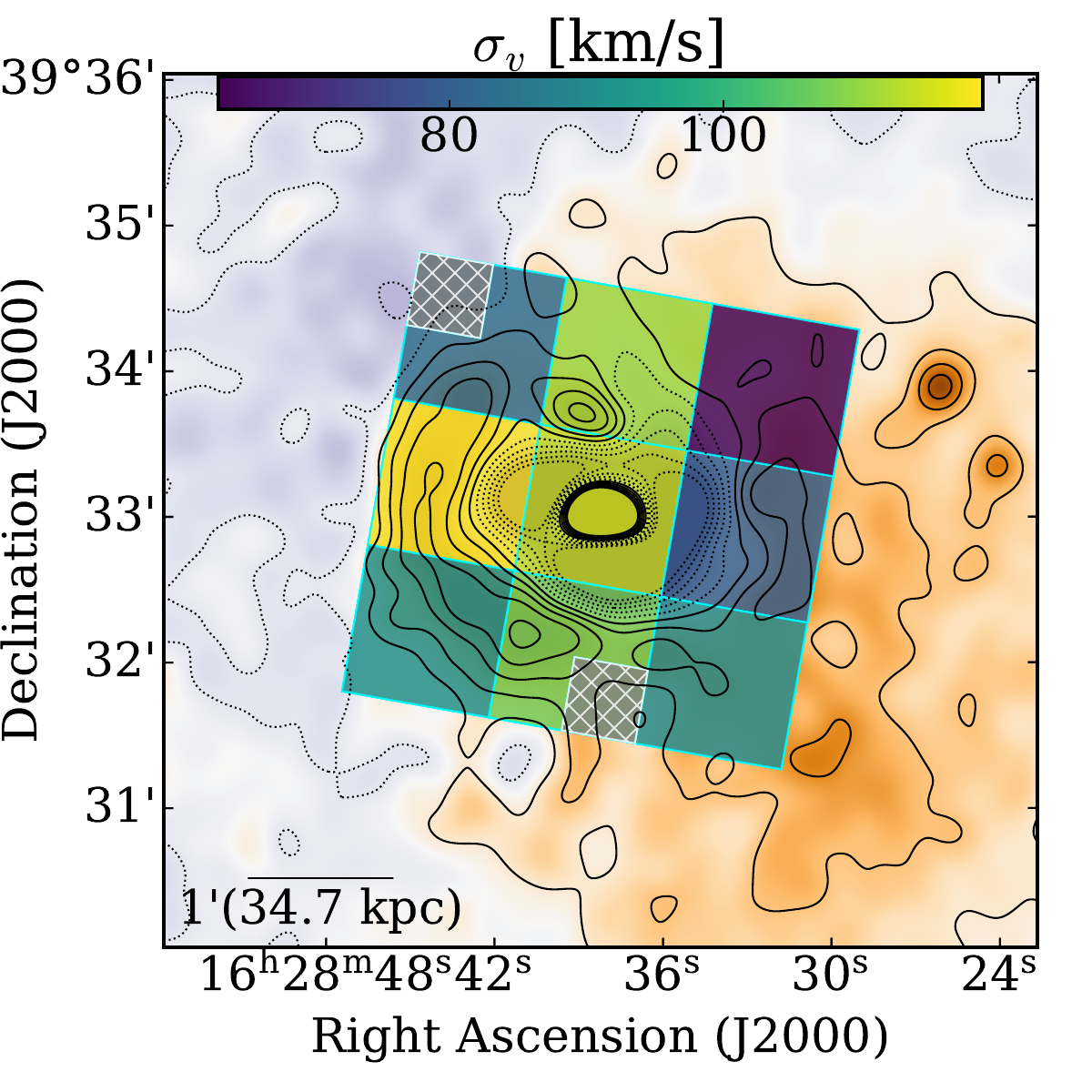}
  \caption{Maps of the best-fit temperature (top left), abundance (top right), bulk velocity relative to the BCG (bottom left), and velocity dispersion (bottom right) derived from the nine-region spatially resolved analysis of the Resolve data.
  {Alt text: Four parameter maps derived from Resolve sub-array analysis are shown using the gnomonic projection in the J2000.0 equatorial coordinates.
  Each map is overlaid on the X-ray residual image with Chandra.
  In each panel, Resolve's field of view is partitioned into nine equal segments, excluding pixels twelve and twenty-seven, with the value of each parameter represented by color thereon.
  North is up, and east is to the left.
  Right ascension is given in sexagesimal format on the abscissa, declination in decimal format on the ordinate.}
}
  \label{fig:4pix_map}
\end{figure*}

\begin{table*}[h]
    \centering
    \begin{threeparttable}   
    \caption{Results of the $2\times2$ pixel ($1'\times1'$) spectral analysis described in Section~\ref{sec:spa-res}.}
    \begin{tabular}{lccccc}
        \toprule
        \multirow{2}{*}{Region} & Temperature & Abundance & $v_{\mathrm{bulk}}$\tnote{*} & $\sigma_{v}$ & C-statistic/d.o.f\\
        & (keV) & (Solar) & (km s$^{-1}$) & (km s$^{-1}$) & \\
        \midrule
        center4pix & $3.33^{+0.05}_{-0.05}$ & $0.70^{+0.03}_{-0.03}$ & $-40^{+9}_{-8}$ & $114^{+10}_{-10}$ & $6430/7942$ \\
        North4pix & $3.69^{+0.08}_{-0.09}$ & $0.62^{+0.03}_{-0.03}$ & $-26^{+12}_{-9}$ & $111^{+13}_{-13}$ & $5001/6148$ \\
        South3pix & $3.82^{+0.09}_{-0.10}$ & $0.63^{+0.04}_{-0.03}$ & $-64^{+11}_{-12}$ & $107^{+15}_{-15}$ & $4259/5268$ \\
        West4pix & $3.68^{+0.08}_{-0.10}$ & $0.56^{+0.03}_{-0.03}$ & $-36^{+10}_{-10}$ & $80^{+15}_{-16}$ & $4659/5848$ \\
        East4pix & $3.72^{+0.08}_{-0.08}$ & $0.57^{+0.03}_{-0.03}$ & $-41^{+12}_{-10}$ & $119^{+12}_{-12}$ & $5080/6297$ \\
        NW4pix & $3.87^{+0.09}_{-0.14}$ & $0.49^{+0.03}_{-0.03}$ & $23^{+14}_{-8}$ & $63^{+18}_{-21}$ & $3632/4493$ \\
        NE3pix & $3.69^{+0.11}_{-0.10}$ & $0.59^{+0.04}_{-0.03}$ & $-25^{+12}_{-11}$ & $83^{+15}_{-16}$ & $3752/4521$ \\
        SW4pix & $4.03^{+0.10}_{-0.12}$ & $0.53^{+0.03}_{-0.03}$ & $-27^{+10}_{-12}$ & $90^{+15}_{-16}$ & $4126/5132$ \\
        SE4pix & $4.10^{+0.12}_{-0.12}$ & $0.52^{+0.03}_{-0.03}$ & $-50^{+11}_{-12}$ & $91^{+15}_{-16}$ & $4093/4885$ \\
        \bottomrule
        \end{tabular}
    \begin{tablenotes}
    \small
    \item[*]  Velocity offset from the BCG ($cz = 9295~\mathrm{km~s}^{-1}$; see Section 3.1). A barycentric correction of $-12.5~\mathrm{km~s}^{-1}$ was applied.
    \end{tablenotes}
    
    \label{tab:params}
    \end{threeparttable}
\end{table*}

\begin{figure*}
    \centering
    \includegraphics[width=0.48\textwidth]{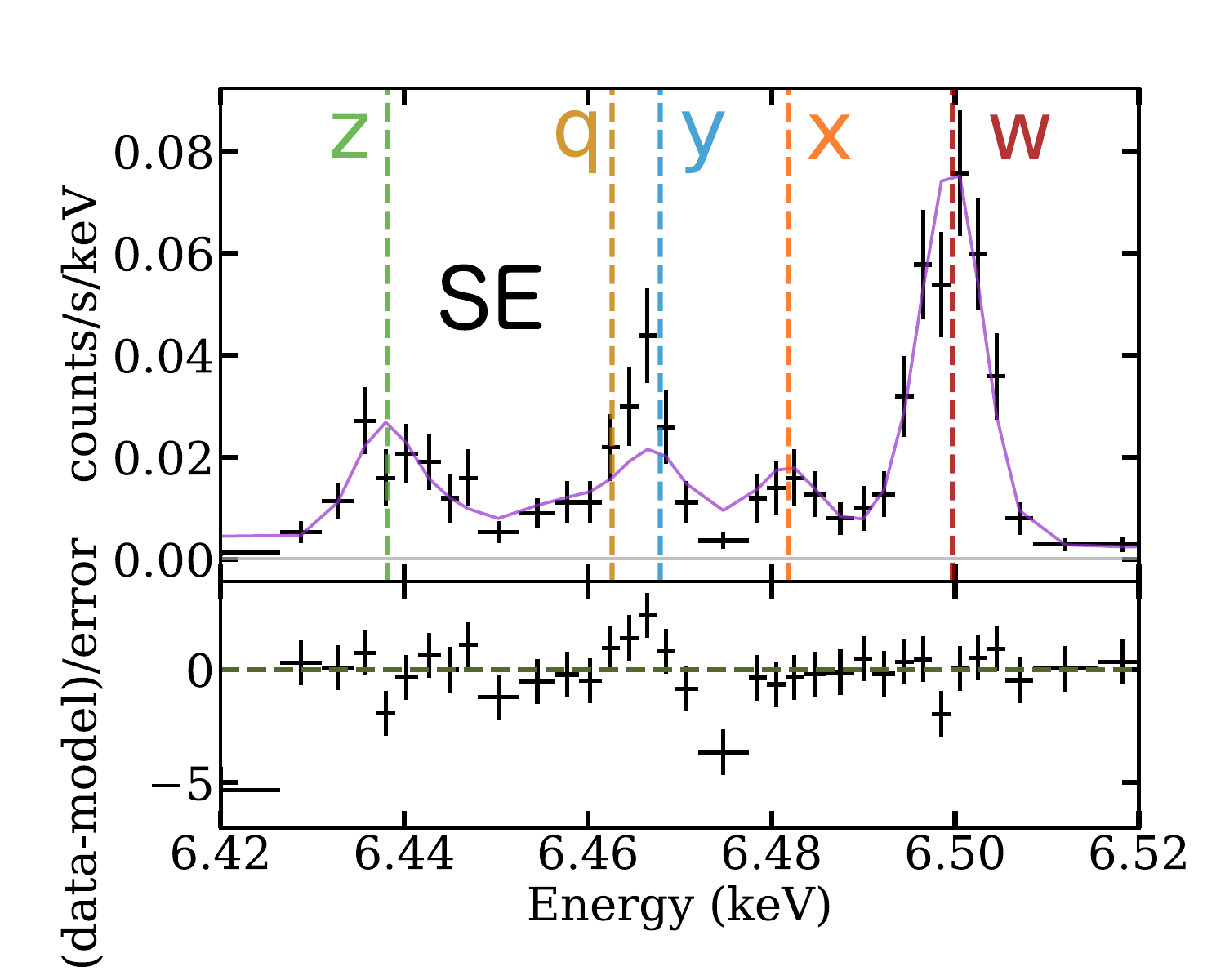}
    \includegraphics[width=0.48\textwidth]{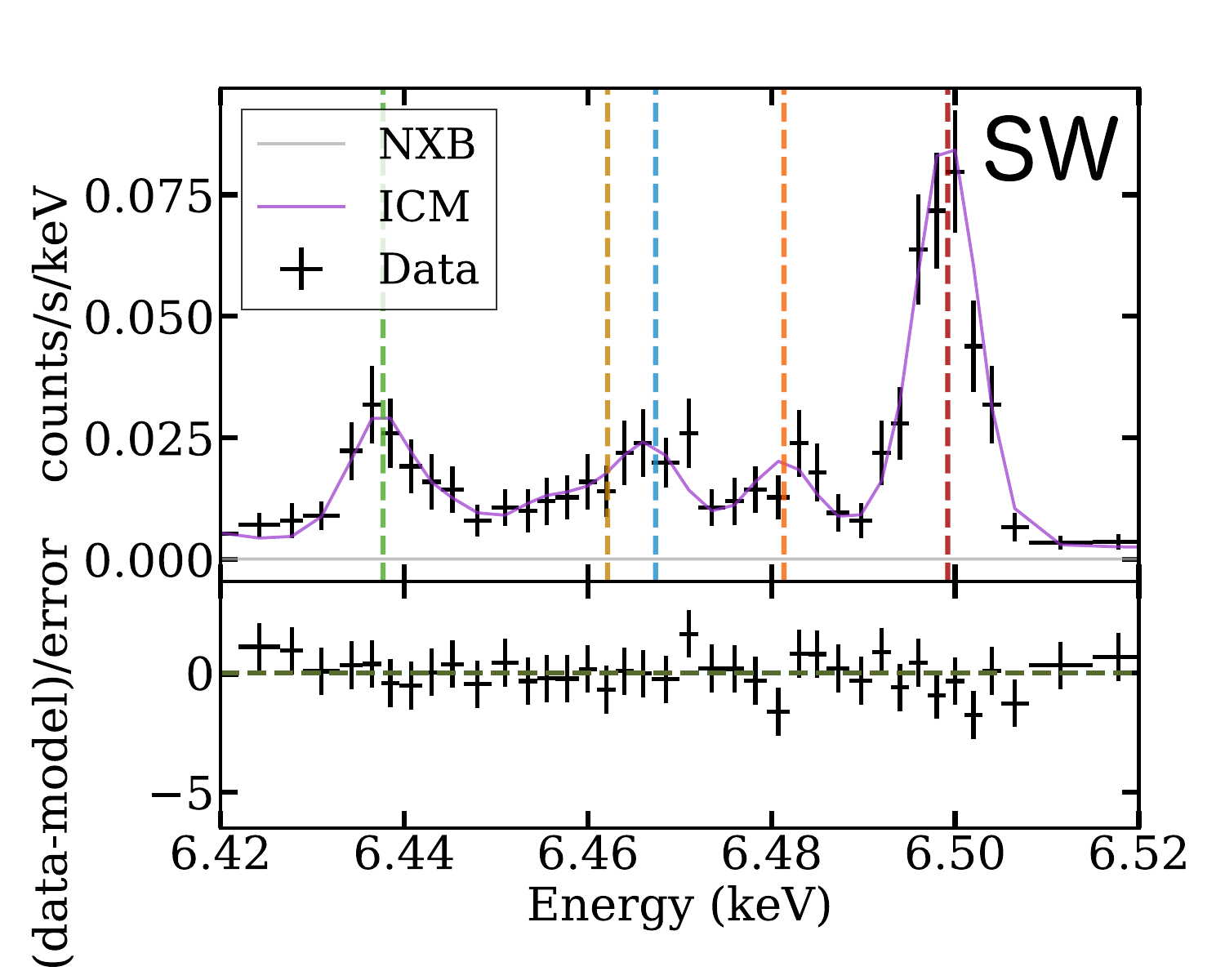}
\caption{The best-fit spectra and residuals for the SE (left) and SW (right) regions obtained from the spatially resolved spectral analysis. The vertical lines indicate the energies of the  $z, q, y, x$, and $w$ lines.
{Alt text: The Resolve spectra in the six-point-four-two-  to six-point-five-two-kiloelectronvolt band, focusing on the iron-helium alpha complex, and best-fit models are shown by data points and model lines. The two representative regions of southeast and southwest are displayed. In each panel, the upper panel shows the data points in units of count per second per kiloelectronvolt, and the residuals are shown in the lower panel, both of which are plotted against photon energy in units of kiloelectronvolt.
}
}
\label{fig:specSSE}
\end{figure*}

\begin{table*}
  \centering
  \begin{threeparttable}
  \caption{Best-fit parameters for the two-temperature model applied to the Resolve FOV-integrated spectrum. }
  \label{tab:fov_fit_table}
  \begin{tabular}{lccccccc}
    \toprule
    \multirow{3}{*}{Model} & \multicolumn{2}{c}{Temperature} & \multicolumn{1}{c}{Abundance} & \multicolumn{1}{c}{Redshift} & \multicolumn{1}{c}{$v_{\mathrm{bulk}}$\tnote{*}} & \multicolumn{1}{c}{$\sigma_{v}$} & \multicolumn{1}{c}{C-statistic/d.o.f} \\
     & \multicolumn{2}{c}{($\mathrm{keV}$)} & \multicolumn{1}{c}{($\mathrm{Solar}$)} & \multicolumn{1}{c}{($10^{-2}$)} & \multicolumn{1}{c}{($\mathrm{km/s}$)} & \multicolumn{1}{c}{($\mathrm{km/s}$)} & \multicolumn{1}{c}{} \\\hline
     \multicolumn{8}{c}{global fit}\\
  2T baseline  & $4.31^{+0.10}_{-0.11}$ & $2.19^{+0.19}_{-0.14}$ & $0.61^{+0.01}_{-0.01}$ & $3.094^{+0.001}_{-0.001}$ & $-30^{+4}_{-3}$ & $100^{+5}_{-5}$ & 13592/14741 \\
    2T (ex-w) & $4.09^{+0.14}_{-0.15}$ & $1.84^{+0.34}_{-0.41}$ & $0.65^{+0.01}_{-0.01}$ & $3.094^{+0.001}_{-0.001}$ & $-32^{+4}_{-3}$ & $102^{+7}_{-7}$ & 13566/14739 \\
    2T (ex-wz) & $3.98^{+0.10}_{-0.08}$ & $1.55^{+0.30}_{-0.26}$ & $0.66^{+0.02}_{-0.02}$ & $3.094^{+0.001}_{-0.001}$ & $-32^{+4}_{-3}$ & $102^{+8}_{-8}$ & 13559/14737 \\
    2T (rsapec) & $4.04^{+0.09}_{-0.11}$ & $1.66^{+0.31}_{-0.34}$ & $0.64^{+0.01}_{-0.01}$ & $3.093^{+0.001}_{-0.001}$ & $-33^{+4}_{-3}$ & $93^{+5}_{-5}$ & 13567/14740 \\
     \multirow{2}{*}{2T-Free} & $4.30^{+0.09}_{-0.13}$ &  & $0.62^{+0.02}_{-0.02}$ & $3.097^{+0.002}_{-0.001}$ & $-24^{+5}_{-4}$ & $84^{+9}_{-7}$ & \multirow{2}{*}{13573/14738} \\
     &  & $2.13^{+0.11}_{-0.25}$ & $0.58^{+0.09}_{-0.08}$ & $3.066^{+0.012}_{-0.019}$ & $-113^{+36}_{-55}$ & $227^{+86}_{-51}$ &  \\\hline
      \multicolumn{8}{c}{local fit}\\
    2T (2.34-2.95 keV) & 4.31\tnote{$\dagger$} & 2.19\tnote{$\dagger$} & $0.63^{+0.03}_{-0.03}$ & $3.079^{+0.008}_{-0.008}$ & $-76^{+24}_{-24}$ & $132^{+50}_{-64}$ & 1129/1209 \\
    2T (2.95-3.12 keV) & 4.31\tnote{$\dagger$} & 2.19\tnote{$\dagger$} & $0.60^{+0.05}_{-0.05}$ & $3.078^{+0.016}_{-0.016}$ & $-79^{+46}_{-46}$ & $207^{+58}_{-56}$ & 342/337 \\
    2T (3.12-3.65 keV) & 4.31\tnote{$\dagger$} & 2.19\tnote{$\dagger$} & $0.55^{+0.04}_{-0.04}$ & $3.088^{+0.016}_{-0.016}$ & $-50^{+46}_{-47}$ & $125^{+75}_{-125}$ & 1071/1055 \\
    2T (3.65-4.3 keV) & 4.31\tnote{$\dagger$} & 2.19\tnote{$\dagger$} & $0.57^{+0.03}_{-0.03}$ & $3.086^{+0.008}_{-0.008}$ & $-56^{+24}_{-24}$ & $102^{+37}_{-43}$ & 1362/1295 \\
    2T (6-7 keV) & 4.31\tnote{$\dagger$} & 2.19\tnote{$\dagger$} & $0.60^{+0.01}_{-0.01}$ & $3.095^{+0.001}_{-0.001}$ & $-29^{+4}_{-3}$ & $100^{+5}_{-5}$ & 1820/1924 \\
    \bottomrule
  \end{tabular}

  \begin{tablenotes}
  \small
  \item[*]  Velocity offset from the BCG ($cz = 9295~\mathrm{km~s}^{-1}$; see Section 3.1). A barycentric correction of $-12.5~\mathrm{km~s}^{-1}$ was applied.
  \item[$\dagger$] Fixed at the temperatures derived from the 2T baseline fit.
  \end{tablenotes}
  \end{threeparttable}
\end{table*}

\subsection{Spatially Resolved Spectral Analysis}
\label{sec:spa-res}

To investigate possible spatial variations,
we divided the Resolve FOV into nine regions of 2$\times$2 pixels ($\sim 1'\times 1'$) and fitted the spectra with a single-temperature (1T) \texttt{bapec} model for the ICM over the 2--12 keV band to provide a first-order assessment of spatial structure.
In this initial analysis, the effects of spatial spectral mixing (SSM), caused by photons originating from adjacent regions, were not taken into account.

The resulting maps of temperature, metal abundance, bulk velocity and velocity dispersion overlaid on the Chandra image are shown in Figure \ref{fig:4pix_map}.
The corresponding best-fit spectral parameters are listed in Table~\ref{tab:params}.
Consistent with previous Chandra studies \citep{johnstone2002, nulsen2013},  the central 1$'\times 1'$ region exhibits a lower temperature and higher metal abundance than the surrounding regions.
In contrast, the bulk velocities are nearly uniform at approximately $-30 ~\mathrm{km\,s^{-1}}$, except for the $1'\times1'$ region toward the northwest. Although the FOV covers the inner portion of the plume-like structure toward the southwest, possibly associated with the sloshing spiral \citep{machado2022}, no statistically significant velocity differences are observed in this direction.
The velocity dispersions are around $100~\mathrm{km\,s^{-1}}$. Although the best-fit values span roughly $63$--$119~\mathrm{km\,s^{-1}}$, the spatial variation is modest once the statistical uncertainties are taken into account. Slightly higher values are found in the central, northern, southern, and eastern regions, but the excess is small.

Given the absence of significant velocity structure across the FOV, the primary results presented in this paper are based on the spectrum extracted from the entire FOV. A more detailed spatially resolved analysis incorporating the SSM effect is currently in preparation and will be presented in a forthcoming paper.

Figure~\ref{fig:specSSE} compares the Fe~\textsc{xxv} He$\alpha$ complex in the SE and SW $1'\times1'$ regions. 
The best-fit model reproduces the overall line structure in the SW region, including the  resonance line ($w$), the intercombination lines ($x$ and $y$), the forbidden line ($z$), and the Fe~\textsc{xxiv} satellite line ($q$) located close to the $y$ line.
In the SE region, the $w$, $x$ and $z$ lines are also broadly consistent with the model, whereas a pronounced excess is seen in the $y$ line band, 
despite similar best-fit temperatures in the two regions.
In the 6.462--6.469 keV interval around the $y$ line, the SE spectrum contains 52 counts, compared with 32 counts expected from the model.

The SE region includes the surface brightness jump identified by the Chandra observation \citep{nulsen2013}.
Although PSF scattering across this local structure may contribute to the observed excess, no comparable excess near the $y$ line is seen in the other $1'\times1'$ regions.
Thus, even allowing for PSF contamination, the excess is likely associated with the SE region.

\subsection{Field-of-View Analysis}

\subsubsection{Full-band fits}
\label{sec:Full-band-fits}
Because Abell~2199 is known to exhibit a multi-temperature structure within the Resolve FOV, we fitted the FOV-integrated spectrum with a baseline two-temperature (2T) \texttt{bapec} model over the 2--12 keV energy range.
The abundance, redshift, and velocity dispersion were tied between the two components.
The best-fit parameters and the C-statistic are listed in Table \ref{tab:fov_fit_table}.
The baseline model yields a C-statistic of 13592 for 14741 degrees of freedom (d.o.f). 
Figure \ref{fig:fov-spectrum} shows the observed spectrum and the best-fit baseline 2T model.
The overall continuum and most emission lines are well reproduced, although small residuals remain around several line complexes, most notably a deficit in the resonance ($w$) line of the Fe He-$\alpha$ triplet.
This fit yields a LOS bulk velocity of $-30~\mathrm{km~ s^{-1}}$  and a velocity dispersion of $\sim100~\mathrm{km ~s^{-1}}$.
The temperatures are 4.3 keV and 2.2 keV, consistent with the projected temperature at 1--1.5\arcmin\ and with the central cool component, respectively, previously identified with Chandra observations \citep{nulsen2013}.

With Hitomi, \citet{hitomicollaboration2018b} detected a similar deficit in the resonance line ($w$) in the Fe He$\alpha$ triplet in the core of the Perseus center and interpreted it as evidence for resonant line scattering, because the optical depth of this line exceeds unity for the measured level of velocity dispersion. 
Resonant scattering can modify the apparent shape of the $w$ line and thus potentially bias the inferred velocity broadening.
To evaluate the impact of the $w$--line deficit on the velocity measurements,
we excluded the $w$--line from the bapec model and instead represented it with an additional Gaussian component. 
The line centroid was fixed at the theoretical rest-frame energy, with its redshift tied to that of the \texttt{bapec} components.
Hereafter, we refer to this model as the 2T (ex-w) model. 
As listed in Table \ref{tab:fov_fit_table},   the best-fit parameters are nearly identical to those obtained with the baseline 2T model; however, the C-statistic improves to 13566 ($\Delta$ C-statistic = 26 for two additional parameters).
Next, to compare the line ratio between the forbidden line ($z$) and the resonance ($w$) lines of Fe He$\alpha$, we excluded these two lines from the bapec model and added two Gaussians.
The central energies were fixed at the theoretical rest-frame values and tied to the same redshift as the ICM components. Although the widths ($\sigma$) of the two Gaussian components were allowed to vary independently, the best-fit values were consistent within the statistical uncertainties,
and the derived velocity parameters also showed no significant changes.
Hereafter, we refer to this model as the  2T (ex-wz) model.

We also tested \texttt{rsapec} model, which accounts for resonant scattering assuming an isothermal plasma 
\citep{chakraborty2023}. In this test, we replaced the two bapec components with two \texttt{rsapec} components and refitted the spectrum using XSPEC  version 12.15.1.
This model reproduces the observed Fe He$\alpha$ triplet reasonably well, 
yielding a C-statistic comparable to that of the 2T (ex-w) model. 
The best-fit bulk velocity is consistent with that obtained from the baseline model, while the inferred velocity dispersion is slightly smaller.
These results suggest that resonance-line suppression, if present, does not significantly bias the velocity measurements.

\begin{figure}[h]
    \includegraphics[width=0.5\textwidth]{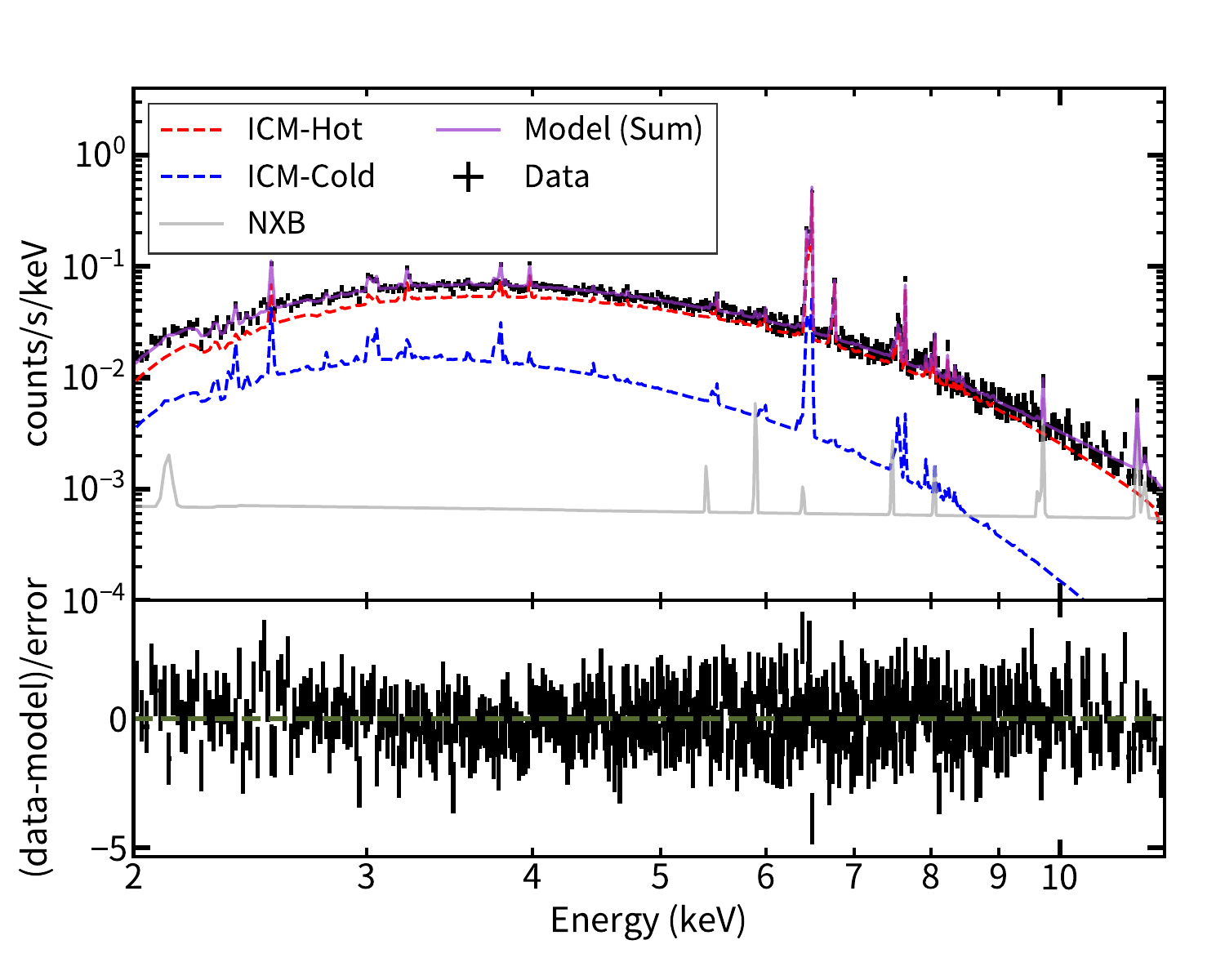}
    \caption{Spectrum and best-fit model of 2T-Free model fitting described in the latter part of Section~\ref{sec:Full-band-fits}. Contributions from the two thermal components (dashed lines) and the NXB (gray solid line) are also shown.
    Alt text: A plot showing the Resolve spectrum and best-fit two-temperature model which allowed the redshift, velocity dispersion and abundance of two components to vary independently. In each panel, the upper section plots data and models in counts per second per kiloelectronvolt versus photon energy, and the lower section shows residuals.}
    \label{fig:2T-Free-spec}
\end{figure}
To investigate the velocity structure of each thermal component, we allowed the redshift, velocity dispersion,  and abundance  of the two components to vary independently and refitted the spectrum. 
We refer to this model as the 2T-Free model. Figure \ref{fig:2T-Free-spec} shows the observed spectrum with the best-fit  2T-Free model. 
While the best-fit temperatures and abundances of the two thermal components and the bulk velocity and velocity dispersion of the hotter component are close to those obtained from the baseline 2T model, the cooler component shows differences of approximately $100~\mathrm{km ~s^{-1}}$ in both bulk velocity and velocity dispersion with the statistical uncertainties of $\sim 50~\mathrm{km ~s^{-1}}$. 

A tendency of the cooler component to exhibit a higher velocity dispersion has been suggested for M87, Centaurus, and Perseus \citep{xrism_m87_aurora, xrism_centaurus, xrism_perseus_coolcomp}. 
However, \citet{xrism_m87_aurora} noted that this trend remains tentative, in particular due to uncertainties on the energy-scale at lower energies.

The XRISM/Resolve team recommends systematic energy-scale uncertainties of $\pm 1~ \mathrm{eV}$ below 5.4 keV and $\pm 0.3~ \mathrm{eV}$ in the 5.4--8.0 keV range \citep{eckart2025}, which have been adopted in other XRISM studies \citep{xrism_m87_aurora, xrism_centaurus_kondo}. 
Assuming energy errors are distributed randomly across the pixels, energy-scale uncertainty also represents a potential systematic Gaussian line broadening.  The velocities of the cooler component tend to be determined by lower-energy lines, whereas those of the hotter component are mainly determined by strong Fe lines. At 2.5 keV, where sulfur lines are present, an uncertainty of $\pm 1~ \mathrm{eV}$ corresponds to an uncertainty of $\pm 120~ \mathrm{km~s^{-1}}$ on the absolute velocity and a potential overestimation of a line width of $200 ~\mathrm{km~s^{-1}}$ by $40~\mathrm{km~s^{-1}}$ . Similarly, at 3.04 keV, near the Ar He$\alpha$ line that shows the maximum velocity inconsistencies (see the next subsection), $\pm 1~ \mathrm{eV}$ corresponds to $\pm 99~ \mathrm{km~s^{-1}}$ and a potential overestimation of a line width of $200~\mathrm{km~s^{-1}}$ by $26~\mathrm{km~s^{-1}}$. With these considerations, the velocity dispersions of the two components are nearly consistent within the combined statistical and systematic uncertainties. 

In addition, by simply considering the energy-scale uncertainty ($\pm1$ eV below 5.4~keV) for the bulk velocity, the two components are consistent within this systematic uncertainty alone.

\subsubsection{Local-band fits}
\label{sec:local-band-fits}
\begin{figure*}[h] 
    \centering
    % (a)
    \includegraphics[width=0.4\textwidth]{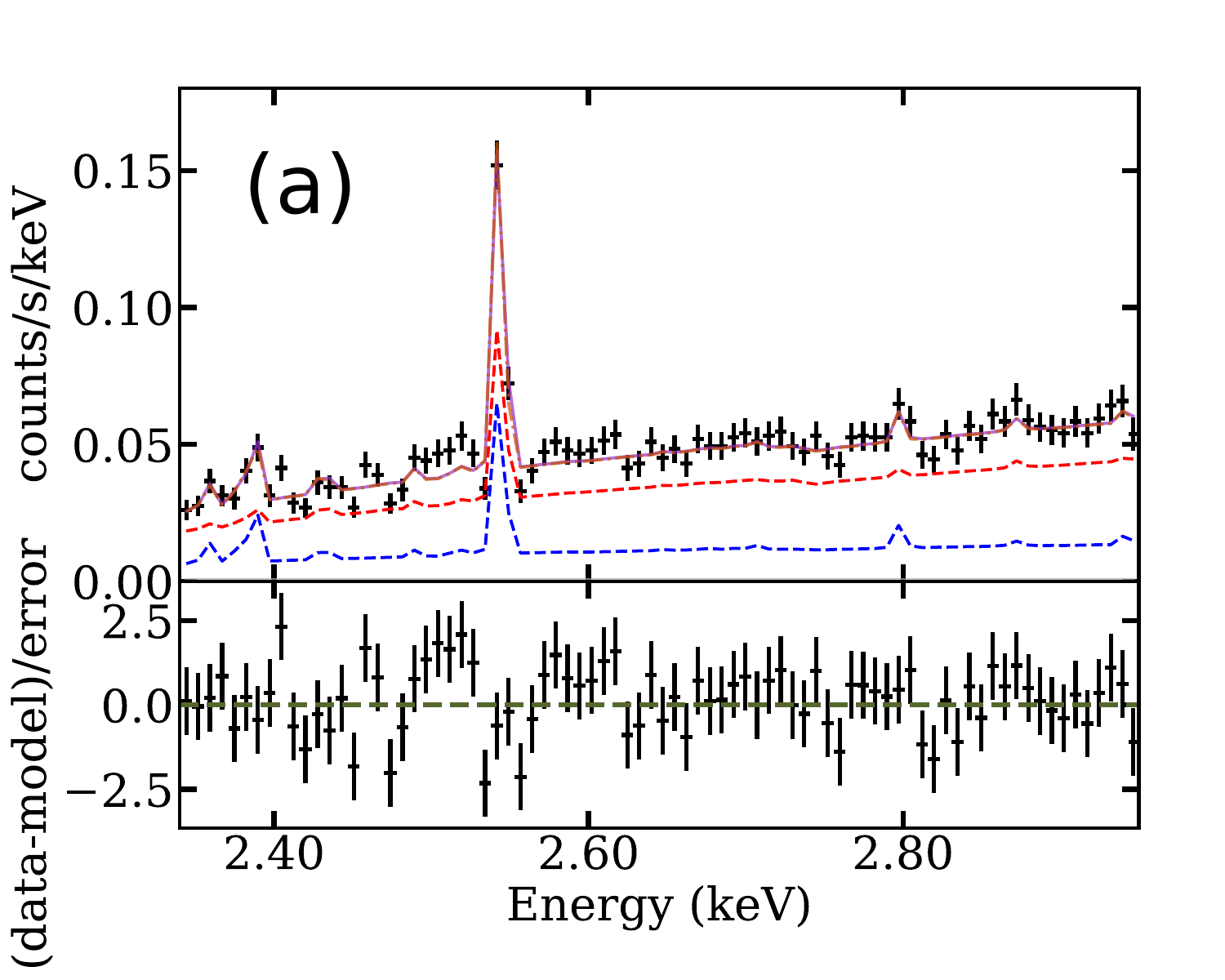}%\hfill
    % (b)
    \includegraphics[width=0.4\textwidth]{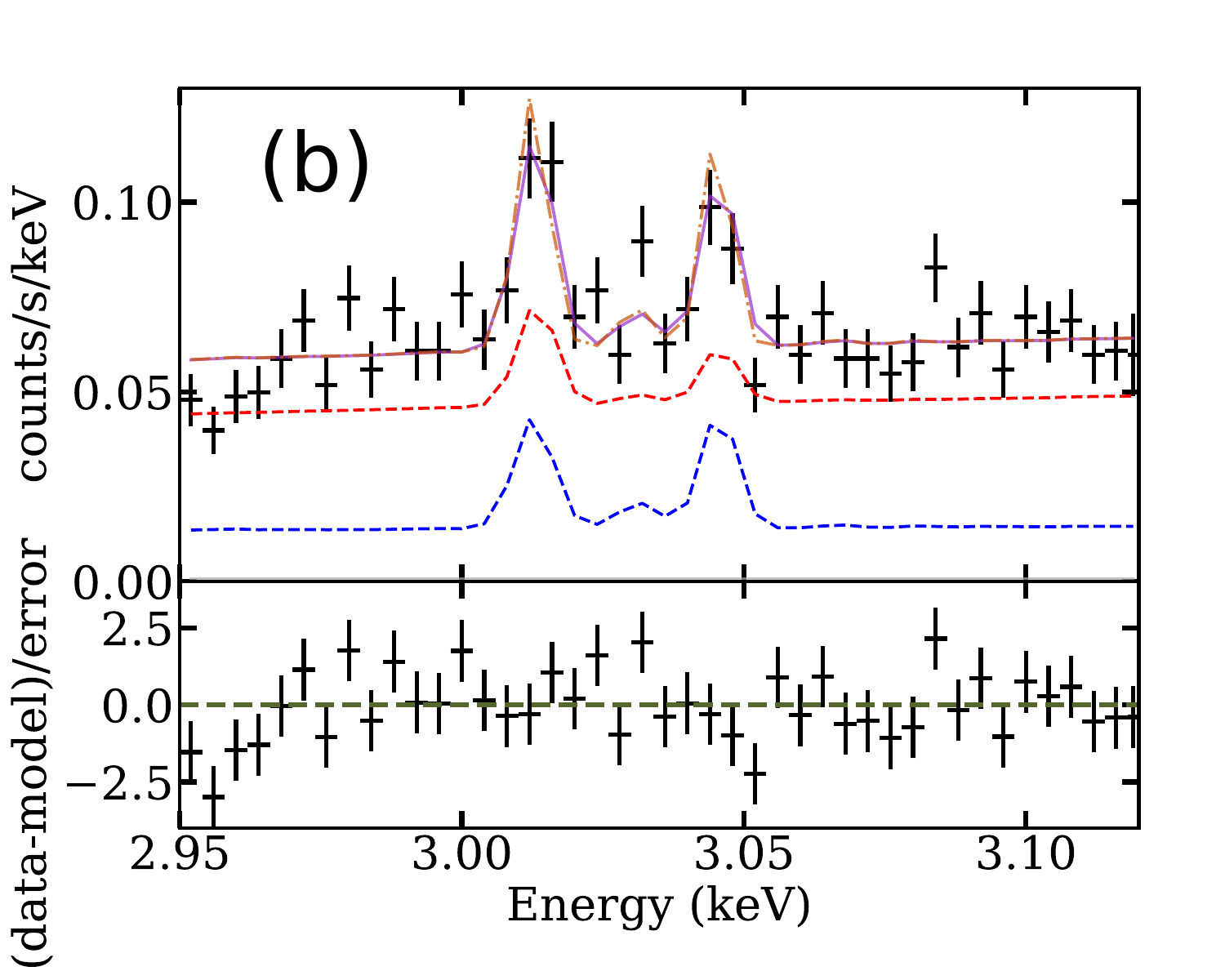}
    
   % \vspace{3mm} 
    
    % (c) 
    \includegraphics[width=0.4\textwidth]{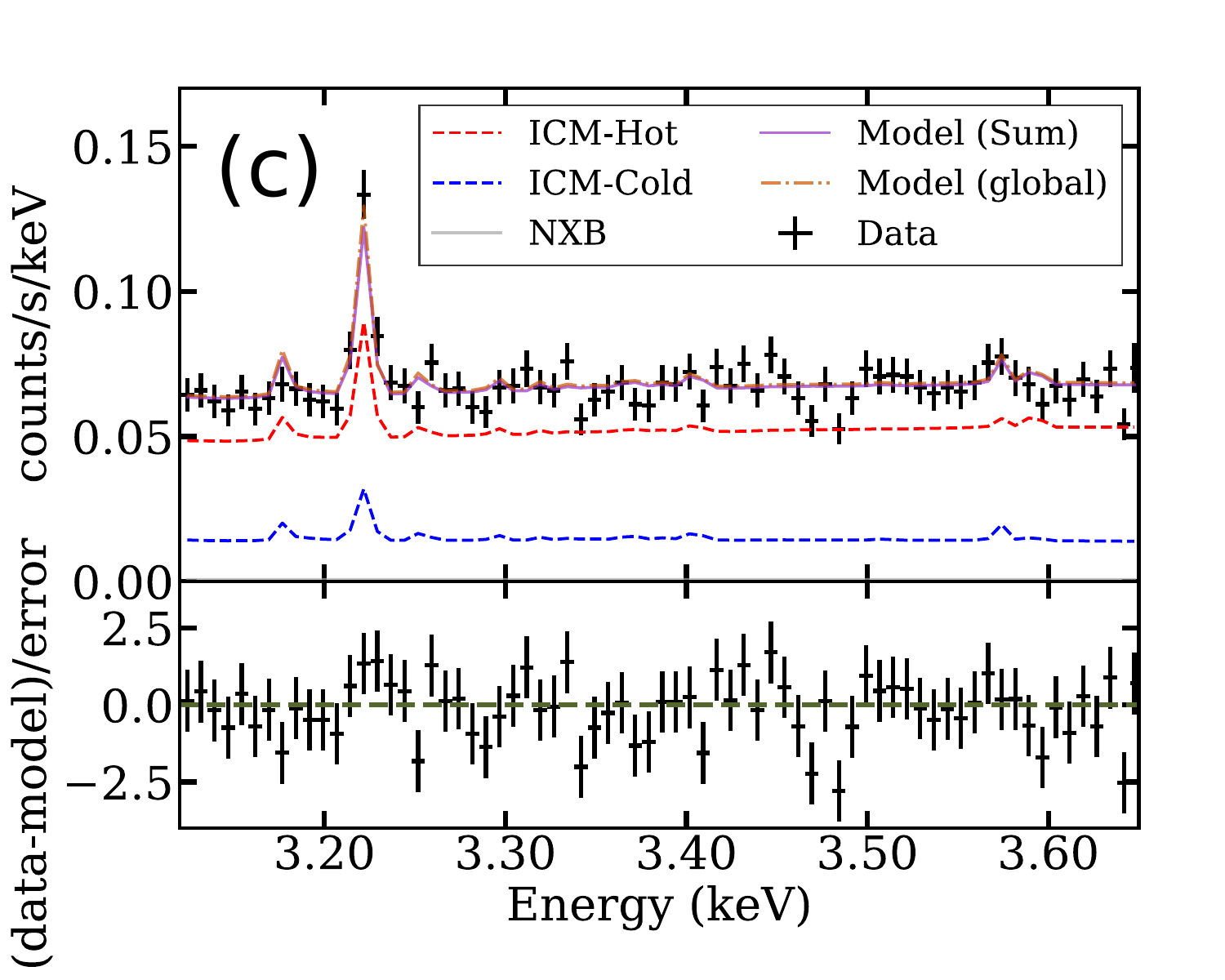}%\hfill
    % (d) 
    \includegraphics[width=0.4\textwidth]{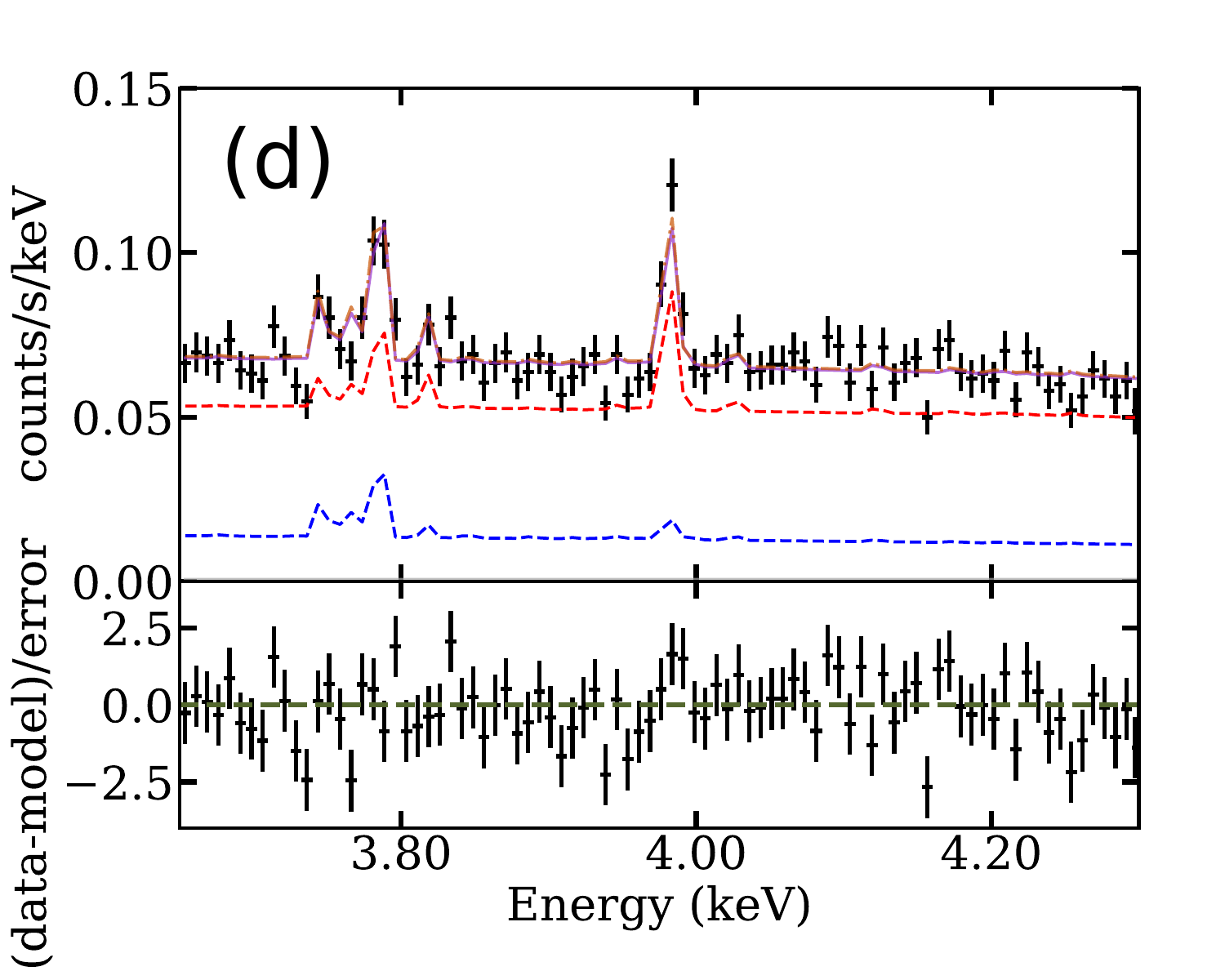}
    
    %\vspace{3mm} 
    
    % (e) 
    \includegraphics[width=0.4\textwidth]{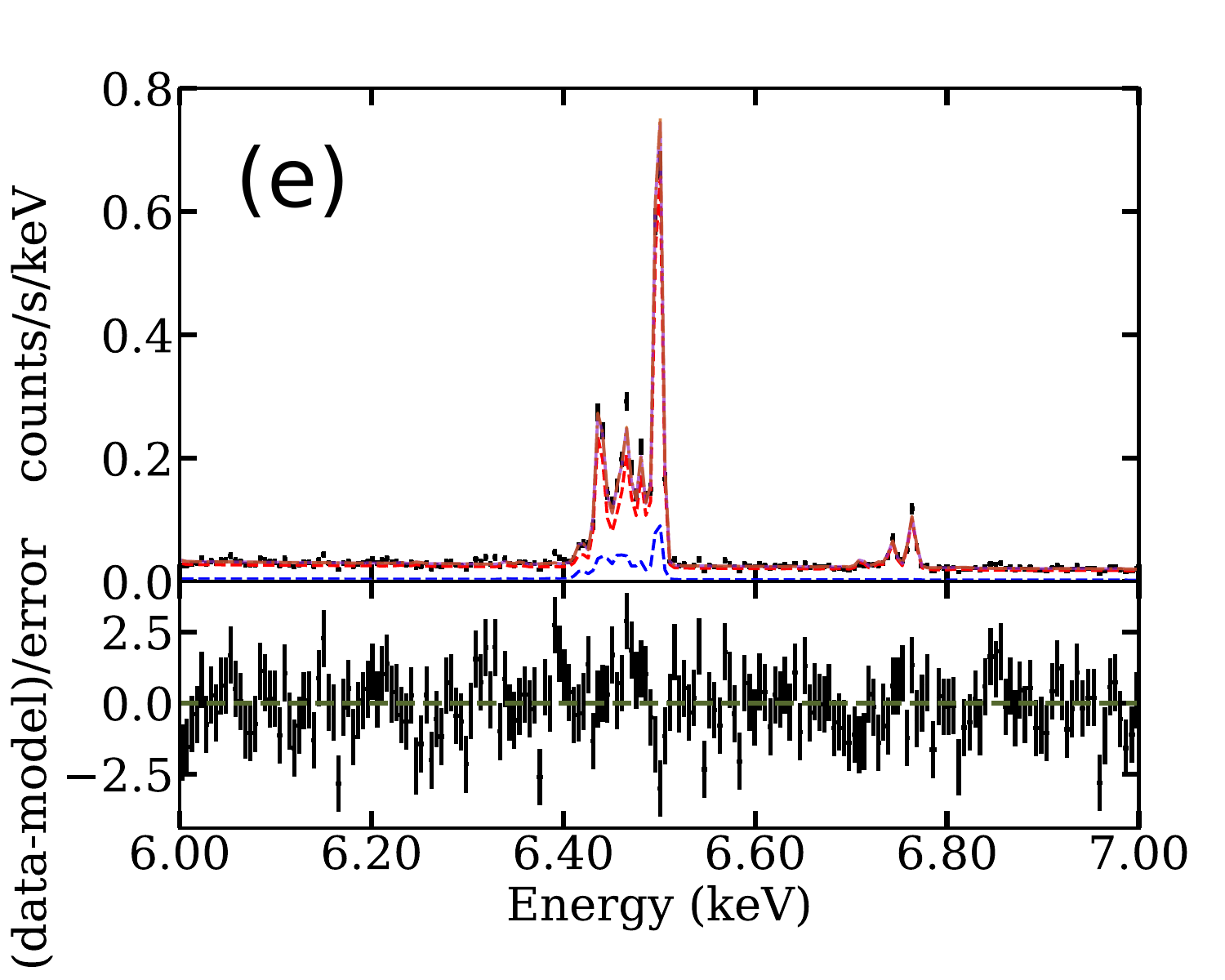}
    
    \caption{Spectra and best-fit models of (a) S band (2.34--2.95 keV), (b) Ar He$\alpha$ band (2.95--3.12 keV), (c) Ar Ly$\alpha$ band (3.12--3.65 keV), (d) Ca band (3.65--4.3 keV), and (e) Fe band (6--7 keV) described in Section~\ref{sec:local-band-fits}. Contributions from the two thermal components (dashed lines) and the NXB (gray solid line) are also shown. For comparison, the best-fit baseline 2T model (orange dash-dot line) is also overlaid on each spectrum. 
    Alt text: A five-panel plot showing the Resolve spectrum and best-fit two-temperature model. In each panel, the upper section plots data and models in counts per second per kiloelectronvolt versus photon energy, and the lower section shows residuals. The five panels focus on specific energy bands and emission lines: sulfur at 2.34 to 2.95 keV (top-left), sulfur and argon at 2.95 to 3.12 keV (top-right), argon at 3.12 to 3.65 keV (middle-left), calcium at 3.65 to 4.3 keV (middle-right), and iron at 6 to 7 keV (bottom).}
    \label{fig:local-spectra}
\end{figure*}

Because the Fe He$\alpha$ triplet dominates the available line statistics, the velocity parameters of both the cool and hot components in the 2T fit are effectively constrained mainly by the Fe He$\alpha$ complex rather than by the weaker lines at lower energies.
To directly measure the velocity parameters and assess their dependence on the thermal components, we performed local-band fits.
We analyze the bands S (2.34--2.95 keV), Ar He$\alpha$ (2.95--3.12 keV), Ar Ly $\alpha$ (3.12--3.65 keV), Ca (3.65--4.3 keV), and Fe
(6--7 keV). 
The temperatures and normalizations  of the two thermal components were fixed at the best-fit values obtained from the baseline 2T model fit, while the remaining parameters were allowed to vary. Figure \ref{fig:local-spectra} shows the observed spectra and the best-fit models for each band. 
The bulk velocities and velocity dispersions derived from the local-band fits are consistent with those obtained from the baseline 2T model within the statistical uncertainties.

The main exception is the 2.95--3.12 keV band, for which the velocity dispersion is higher by about 2$\sigma$ but remains consistent with the baseline 2T model within the combined uncertainty including the systematic uncertainty described in the previous subsection.
The agreement among the different bands indicates that the measured velocity parameters are not strongly dependent on the choice of lines.

\section{Discussion}
\subsection{Implications of the small core bulk velocity}

Sloshing motions are expected to drive bulk flows and turbulence in cool-core clusters, although observed LOS velocity depends on the viewing geometry.

\citet{nulsen2013} suggested that the observed X-ray structure resembles sloshing simulations viewed approximately edge-on. 
Based on numerical simulations, \citet{machado2022} suggested that the sloshing in A2199 is viewed nearly edge-on, with the orbital plane inclined by
 $\sim 70^{\circ}$ to the plane of the sky.
In their best-fit model, a perturber with a mass ratio of 20 to 1 approached from the southwest, passed near the BCG about 0.8~Gyr ago, and is currently located about 2~Mpc northeast of the cluster center.  
In this scenario, the dark-matter density peak of the main cluster is first pulled toward the incoming subcluster, and later bends and drifts in the opposite direction.
This geometry is broadly consistent with the southwest plume-like excess seen in the \textit{Chandra} residual image, which extends beyond the Resolve FOV and likely traces part of the outer sloshing spiral.

Although the model by \citet{machado2022} does not provide a predicted LOS velocity field, 
this configuration suggests that gas bulk motions should occur along the sloshing spiral, particularly near the cold front, while turbulence may be generated beneath the cold front (e.g. \citealt{zuhone2011}).

Our XRISM observation of the A2199 core yields a small velocity difference of approximately 
$30 \ \mathrm{km ~ s^{-1}}$ relative to the BCG, with modest spatial variation across the $104 \ \mathrm{kpc} \times 104 \ \mathrm{kpc}$  Resolve FOV. 
Accounting for the uncertainty in the optical redshift of the BCG, the ICM bulk velocity is consistent with no significant motion relative to the BCG. 
Together with the low velocity dispersion ($\sim100~\mathrm{km s^{-1}}$), this result indicates that the ICM in the inner $\sim100$ kpc of Abell~2199 is dynamically quiescent in the LOS direction. 

The small velocity dispersion disfavors a scenario in which the near-zero bulk velocity is produced mainly by cancelation between unresolved red- and blueshifted flows, because substantial LOS shear would also increase the line width.
This internal quiescence does not necessarily imply that the core is at rest with respect to the cluster as a whole.
Rather, the BCG and the gas within the XRISM field appear to form a kinematically coherent core system that is offset from the mean cluster redshift  by about $200~\mathrm{km~s^{-1}}$ \citep{Lauer2014}.
This offset may reflect the motion of the core within the cluster potential, while the internal velocity structure of the core remains limited.

Several scenarios may explain the observed kinematics. 
The perturbation may have been relatively weak, or the sloshing amplitude in the innermost core may have diminished by the time of our observation.

Alternatively, 
a rotating sloshing spiral may still persist.
The largest LOS velocity contrasts may occur in the outer spiral structure rather than in the central region covered by the current FOV. 
The observed LOS velocities may also depend on the viewing geometry and oscillation phase.

An encounter with a small impact parameter may generate cold fronts without producing a prominent large-scale spiral velocity pattern.

However, such a scenario could disrupt the characteristic central radio structure of 3C 338, making it difficult to maintain its observed morphology, as discussed by \citet{nulsen2013} and \citet{antas2024}.

Future XRISM observations targeting the southwestern plume outside the current Resolve FOV will therefore provide an important test by probing whether larger LOS velocity contrasts emerge along the outer sloshing arm.

A useful comparison is provided by the Centaurus cluster, which is approximately three times closer than A2199 and lacks a bright central AGN.
The Resolve observation of the Centaurus 
reveals substantially larger blueshifted bulk velocities of 130--310~$\mathrm{km ~s^{-1}}$, including a $\sim300~\mathrm{km~ s^{-1}}$ strong blueshifted region at $\sim1\arcmin$ west of the BCG \citep{xrism_centaurus}.
Despite sampling a smaller physical region, the Centaurus core shows a much stronger LOS bulk motion.
This contrast suggests that the difference is not simply a matter of spatial scale, but reflects different merger geometries and dynamical states.
In Centaurus, the Cen 45 subgroup is a plausible perturber and has an optical
velocity offset of $\sim1500~\mathrm{km~ s^{-1}}$ relative to the main Centaurus component.
\citet{walker2013} found enhanced temperature, metallicity, and surface brightness around Cen~45 with XMM-Newton, indicating ongoing interaction between the subgroup and the main cluster.
The large redshift difference between Cen~45 and the main cluster suggests that the perturber motion has a substantial LOS component, which could naturally explain the large bulk-velocity offset observed in the Centaurus core.

\subsection{Velocity Dispersion: Energy balance between turbulent heating and radiative cooling}

A2199 is a typical cool core cluster and hosts a central AGN associated with the bipolar radio lobes extending toward the east and west \citep{burns1983}.
If the velocity dispersion of $\sim100~\mathrm{km~ s^{-1}}$ obtained in this study is attributed to turbulence, 
the dissipation of turbulence can contribute to heating of the ICM.
Following \citet{zhuravleva2014}, 
 the turbulent dissipation rate per unit volume, $Q_{\text{turb}}$, can be estimated as 
\begin{equation}
Q_{\text{turb}} \sim 5\frac{\rho v_{\text{turb}}^3}{l_{t}}
\end{equation}
where $\rho$ is the gas mass density, $v_{\text{turb}}$ is the characteristic (one-dimensional) turbulent velocity amplitude, and $l_t$ is the length scale of the turbulence. 
The radiative cooling rate is expressed as
\begin{equation}
Q_{\text{cool}} = n_e n_i \Lambda(T)
\end{equation}
where $n_e$ and $n_i$ are the electron and ion number densities, respectively, and $\Lambda(T)$ is the temperature-dependent cooling function. 

For Abell~2199, we adopt $v_{\text{turb}} = 100 ~\mathrm{km\ s^{-1}}$  based on the velocity dispersion measured within the Resolve FOV.
 The electron density profile $n_e$ is taken from the deprojected Chandra analysis by \citet{johnstone2002}, which  assumes spherical symmetry.
 We evaluate the cooling function $\Lambda(T)$ using \citet{sutherland1993} for a 4.3 \color{black} keV plasma with solar metallicity.

For $l_t$, we first adopt an emission-weighted LOS depth, defined as
the path length that contributes to 50\% to the integral of $n_e^2$ along the line of sight, following the approach used in the XRISM analysis of the Perseus cluster \citep{xrism_perseus}.
This estimate assumes a relatively large driving scale associated with the cluster-scale sloshing structure.
Because the measured velocity dispersion represents an emission-weighted average over the Resolve FOV, we evaluate both $Q_{\text{turb}}$ and $Q_{\text{cool}}$ at three representative projected radii from the cluster center: 0.5, 1.0, and 1.5 arcmin.  
For example, at $0.5~\mathrm{arcmin}$,  we obtain $l_t$ $\sim 31~\mathrm{kpc}$. 
At all three radii, we obtain  $Q_{\text{turb}} / Q_{\text{cool}}\approx0.1\color{black}$. 

In Abell~2199, 
the multiwavelength morphology around the BCG 
also supports a possible contribution from AGN-driven gas motions. 
As seen in the LOFAR \citep{van-Haarlem2013}  image, 
the bipolar radio lobes are elongated along the east-west direction  
extending to at least 1.1$'$ from the center, well within  the Resolve FOV (Figure \ref{fig:fov-chandra}). 
The BCG also hosts a compact optical emission-line nebula seen in the SDSS/MaNGA data. 
The warm ionized gas is traced to $\sim$ 4.5 kpc radius, entirely within the $1'\times1'$ region, and is elongated in the east-west direction.
It shows a velocity gradient of $\sim$ 200 $\mathrm{km~s^{-1}}$, with
the eastern side blueshifted and the western side redshifted.
If the warm ionized gas is uplifted by the radio AGN, this velocity pattern may indicate that the eastern jet/lobe is approaching and the western jet/lobe is receding along the line of sight.
Such AGN-driven motions may also affect the kinematics of the surrounding hot gas. 

Therefore, we also estimate $Q_{\text{turb}}$, taking the turbulence injection scale to be comparable to the characteristic size of the radio lobes, as adopted in the XRISM Perseus analysis \citep{xrism_perseus}.
For $l_t=4\text{--}23~\mathrm{kpc}$, corresponding to the
 inner and outer radio lobe sizes reported by  \citet{nulsen2013},
 we obtain $Q_{\text{turb}} / Q_{\text{cool}}$ ratios $\approx$0.3 \color{black}  at $l_t=4~\mathrm{kpc}$ and $\approx$0.2 \color{black} at $l_t=23~\mathrm{kpc}$, using the local density appropriate for each scale.
 If we instead adopt the velocity dispersion of $115~\mathrm{km~s^{-1}}$ measured in the central four pixels without SSM analysis, the corresponding ratios increase to $\approx0.5 \color{black}$ and 0.3\color{black}. 
 The dominant uncertainty is the driving scale $l_t$.
 In addition, since $Q_{\text{turb}} \propto v_{\text{turb}}^3$, modest spatial variations in the turbulent velocity can lead to significant systematic uncertainties in the heating estimate. 
Thus, the present calculation should be regarded as an order-of-magnitude estimate rather than a precise energy-balance measurement. 

Despite these uncertainties, the estimated turbulent heating rate remains within the same order of magnitude as the radiative cooling rate for plausible choices of $l_t$.
This suggests that turbulent dissipation is a non-negligible component of the heating budget in the core of Abell~2199.
The presence of both inner and outer radio lobes, together with the result of \citet{antas2024} that the morphology of 3C~338 requires intermittent AGN activity, suggests that the energy input from AGN feedback is episodic.
Although the time-averaged heating efficiency depends on the AGN duty cycle, such episodic feedback may help offset radiative cooling over long timescales.
A more detailed analysis will be presented in a forthcoming paper.

 \subsection{Comparison of ICM velocity dispersions around BCGs observed with XRISM }

\begin{figure*}
  \includegraphics[page=1, width=0.48\textwidth]{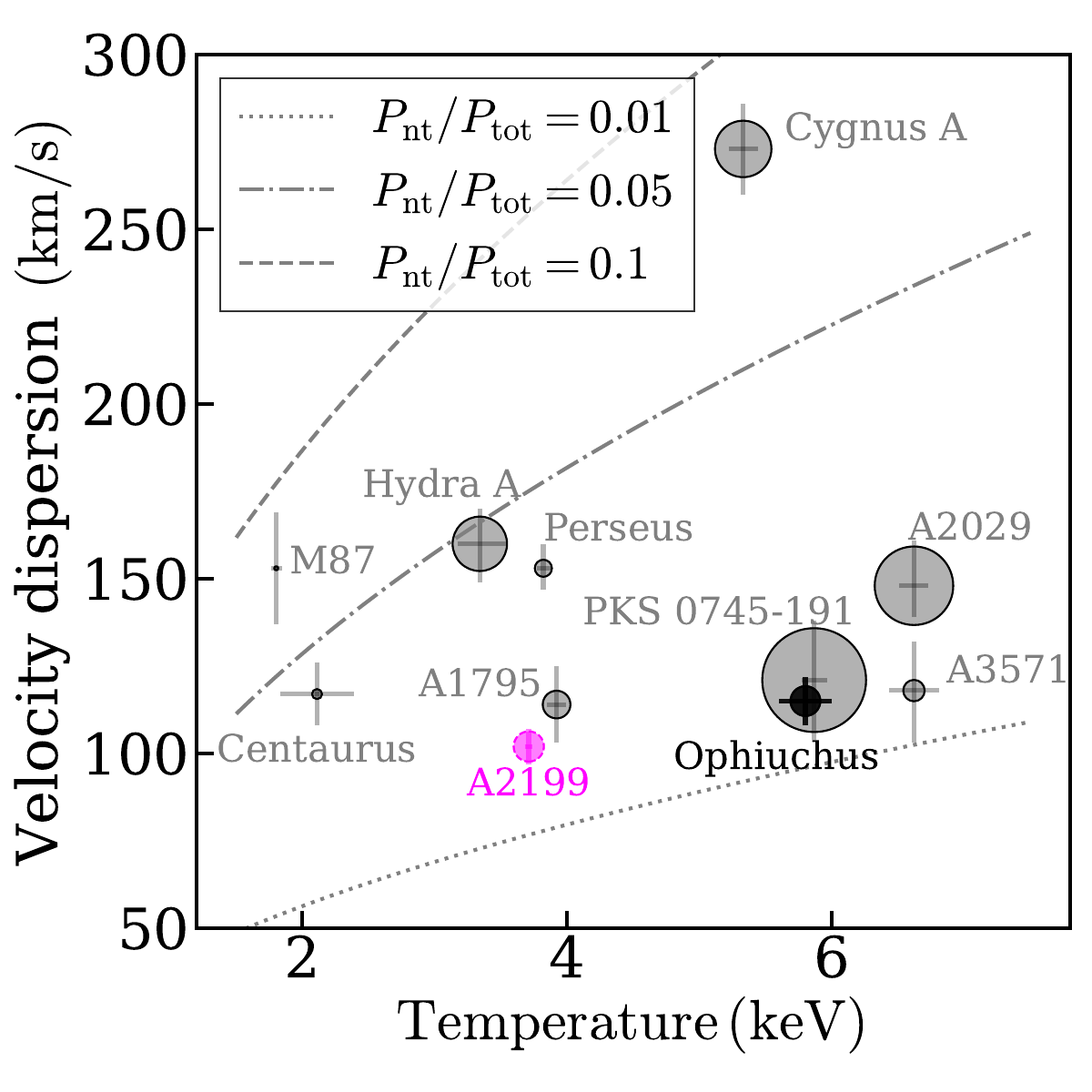}\includegraphics[page=2, width=0.48\textwidth]{vdis-vs-something_plot_all.pdf}
  \includegraphics[page=3, width=0.48\textwidth]{vdis-vs-something_plot_all.pdf}\includegraphics[page=4, width=0.48\textwidth]{vdis-vs-something_plot_all.pdf}
  \caption{
 The velocity dispersions of Abell~2199 and other clusters in the cluster core regions  around the BCGs, plotted against the ICM temperature, the classical cooling flow rate, central radio luminosity, and power of cavities of radio bubbles.  
 Marker areas are proportional to the physical areas of the spectral extraction regions (e.g., the marker for A2199 corresponds to $104 \times 104 \text{ kpc}^2$). \color{black}
 [References] Velocity Dispersion \& Temperature: \citet{xrism_cosmosim} for A2029, Centaurus, M87, Hydra-A, Perseus, PKS 0745-191 and Ophiuchus; \citet{xrism_cygnus} for Cygnus-A; R1 region of \citet{mccall2026_a3571} for A3571; Region 1 of \citet{sarkar2026_a1795} for A1795/ $\dot{M}_{\mathrm{cool}}$: \citet{peres1998} /
 $L_{1.4}$: \citet{birzan2012} /
 $P_{\mathrm{cav}}$: \citet{russell2026} for Ophiuchus; \citet{rafferty2006} for the rest. 
 Alt text: Four-panel plot of galaxy cluster velocity dispersion (y-axis) versus temperature (top-left), classical cooling flow rate (top-right), radio luminosity (bottom-left), and radio bubble cavity power (bottom-right). The top-left panel includes three lines (dotted, dashed, dash-dot) showing different non-thermal fractions.}
  \label{fig:vturb}
\end{figure*}

XRISM observations of cluster cores have shown that velocity dispersions and the associated non-thermal pressure fractions typically lie near the lower end of the range predicted by numerical simulations \citep{xrism_cosmosim}.
In this context, we estimate the Mach number and the corresponding non-thermal pressure fraction in the core of Abell~2199 using the velocity dispersion measured with Resolve.
Because the hotter component dominates the Fe He$\alpha$ triplet, we adopt its temperature to estimate the Mach number.
For a temperature of 4.3 keV, the corresponding sound speed is $c_\mathrm{s}=1071~ \mathrm{km\,s^{-1}}$.
Adopting the velocity dispersion of $100 ~\mathrm{km\,s^{-1}}$, 
and assuming isotropic gas motions, the inferred three-dimensional Mach number is 
$\mathcal{M}_{\mathrm{3D}}=\sqrt{3}\sigma_{\mathrm{v}}/c_{\mathrm{s}}=0.16$. This corresponds to a non-thermal pressure fraction of
$\frac{P_{\mathrm{NT}}}{P_{\mathrm{tot}}}=\frac{\mathcal{M}^2_{\mathrm{3D}}}{\mathcal{M}^2_{\mathrm{3D}}+3/\gamma}=1.4\pm0.2$\%.
We neglect the contribution from the bulk velocity ($\sim30~\mathrm{km\,   s^{-1}}$), since the non-thermal pressure scales with the square of the velocity and the bulk term is therefore subdominant compared to the velocity dispersion.
The top\, left panel of Figure \ref{fig:vturb} compares the observed velocity dispersions in the core regions of eleven  clusters observed with XRISM, as a function of the ICM temperature.
  
 To investigate the possible contributions of sloshing and AGN feedback to ICM gas motions in the vicinity of BCGs, 
 we restricted the comparison sample to cool-core and weak cool-core clusters in which the BCG is located close to the central X-ray surface-brightness peak. 
The non-thermal pressure fraction of Abell~2199 is among the lowest in the current sample, although the spatial scales are not identical among the observations.  

The high velocity dispersion of Cygnus-A \citep{xrism_cygnus} and the central velocity dispersion peaks observed in M87 and Perseus \citep{xrism_m87,xrism_perseus} have been interpreted as signatures of gas motions driven by the central AGN,  possibly superposed on   larger-scale sloshing induced  motions. 
Motivated by this picture, we compared the observed velocity dispersions with the classical cooling-flow rates (top right panel of Figure \ref{fig:vturb}; estimated by \citet{peres1998} in the absence of heating), since AGN feedback is widely considered the primary heating mechanism in cool cores.
Although the classical cooling-flow rates of A2199, Hydra~A, and Perseus span a range of $100-600~M_\odot {\rm yr^{-1}}$, the ratios $Q_{\text{turb}} / Q_{\text{cool}}$ are comparable among these three clusters. 

We also examined possible correlations with the central 1.4 GHz radio luminosity (bottom left panel of Figure \ref{fig:vturb})  and cavity power (bottom right) , using 1.4 GHz radio luminosities from \citet{birzan2012}, and the time-averaged cavity power calculated by \citet{russell2026} for Ophiuchus and \citet{rafferty2006} for the rest. 
For consistency, we adopted the \citet{rafferty2006} value for 
Cygnus~A, although  \citet{wilson2006} reported a cavity power about an order of magnitude larger. However, this difference does not significantly affect our discussion.

 A3571 is excluded from the cavity power comparison because no reliable cavity power estimate is available.

At lower radio luminosities, the velocity dispersion shows no clear trend with variations in radio luminosity, remaining distributed around $100\text{--}150\text{ km s}^{-1}$. Clusters with relatively high radio luminosities—particularly Cygnus~A, Hydra~A, and Perseus—exhibit slightly higher velocity dispersions, resembling the relation between the non-thermal energy fraction and AGN X-ray luminosity reported in \citet{yamada2026}. Note, however, that Cygnus~A hosts the most powerful radio source in the sample, and its large velocity dispersion has been suggested by numerical simulations in \citet{shirotori2026} to originate from the unresolved bulk expansion motion of its cocoon. Excluding Cygnus~A, no clear correlation is found between velocity dispersion and radio luminosity, nor is any correlation found with cavity power.

The lack of a simple trend may arise because the measured gas motions are a mixture of AGN-driven motions and sloshing induced flows: AGN activity can perturb the ICM locally, whereas the volume-averaged velocity dispersion measured over $\sim100$ kpc scales does not scale simply with instantaneous AGN power.
Although the radio luminosity and jet power of Abell~2199 are comparable to, or even higher than, those of the Perseus cluster, its velocity dispersion is substantially lower.
The Centaurus cluster exhibits the weakest indicators of AGN activity, yet its velocity dispersion is slightly higher than that of Abell~2199; in this system,
 gas motions in the core may have a substantial contribution from sloshing 
 \citep{xrism_centaurus}. 
Thus, despite its plume-like structure possibly associated with sloshing and its relatively high AGN power, Abell~2199 shows one of the lowest velocity dispersions in the current sample.

Our results are consistent with \citet{mcnamara2026}, who similarly reported no clear trend between velocity dispersion and jet power for a sample of clusters observed with XRISM (Note that their sample includes Coma which are not included in our plots). 
Previous imaging study that estimated turbulent velocities from X-ray surface brightness fluctuations across multiple clusters also support this lack of correlation in relaxed systems. While \citet{eckert2017} who use XMM-Newton, Chandra, and ROSAT/PSPC data, found a positive correlation between radio power and turbulent velocity for clusters hosting radio halos (i.e., relatively merging systems), \citet{dupourque2024} found no such correlation in their sample observed with XMM-Newton, which omits less-relaxed clusters. 
The absence of a clear trend may reflect the limited sample size and possible time lags between AGN power and the ICM velocity field.
A larger XRISM sample will be crucial for establishing the dominant drivers of gas motions in cluster cores.

\subsection{Optical depth effects on the Fe He-$\alpha$ resonance line}
\begin{figure}[htbp]
\includegraphics[width=0.48\textwidth]{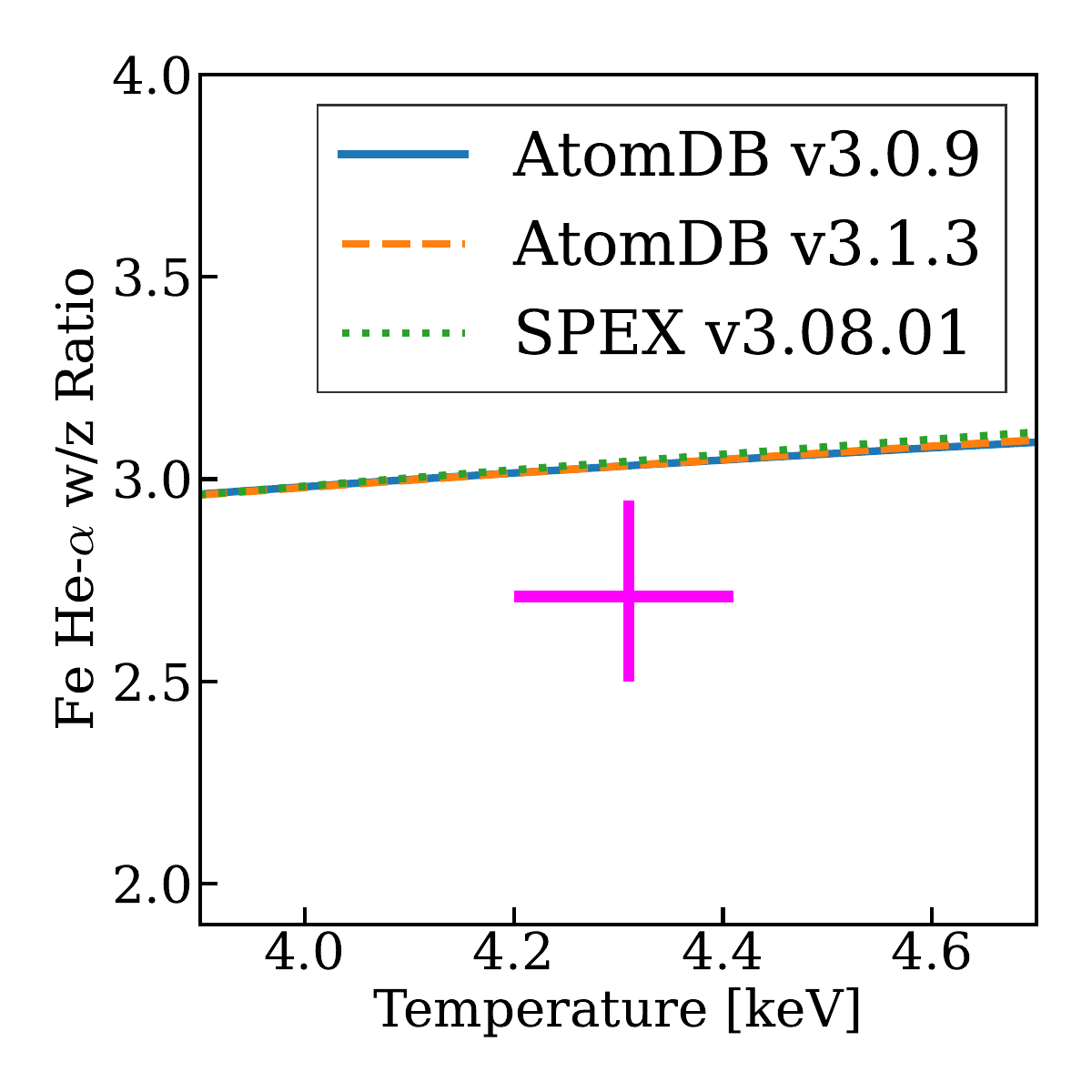}
\caption{The best-fit $w$-to-$z$ ratio obtained with the 2T (ex-wz) model, and theoretical expectations with AtomDB v3.0.9 (solid line),
AtomDB v3.1.3 (dashed line) and SPEX v3.08.1 (dotted line) plotted as a function of  temperature.
{Alt text: One data point and three lines are plotted on a single graph with temperature on the abscissa and flux ratio on the ordinate to compare the observed flux and model predictions. The temperature range is from four to four point seven kiloelectronvolts.}
}
\label{fig:wzratio}
\end{figure}

\begin{figure}[htbp]
    \centering
    \includegraphics[width=0.48\textwidth]{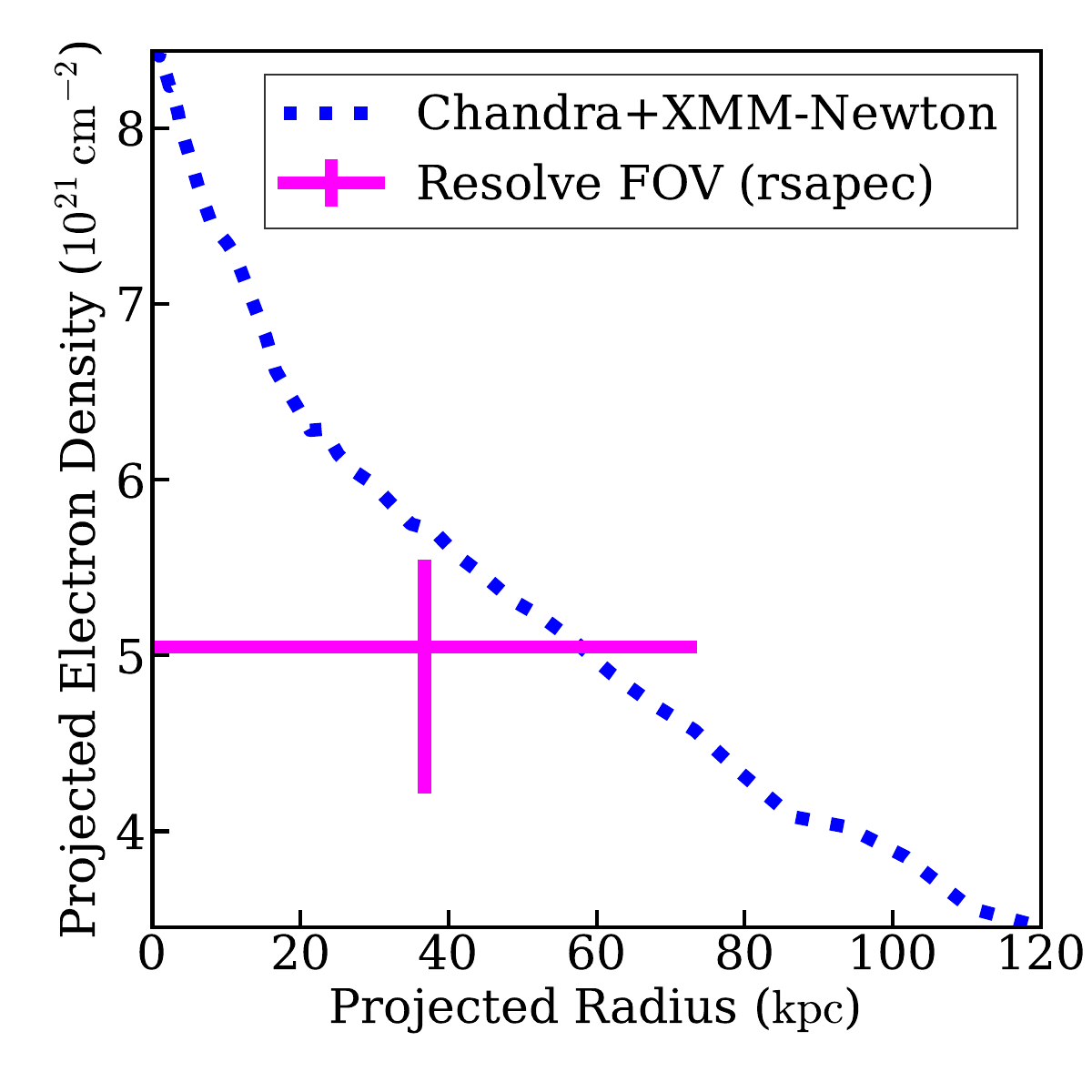}
    \caption{Projected electron column density profile derived from Chandra \citep{johnstone2002} and XMM-Newton \citep{Mirakhor2020} data (dashed line) and that from the \texttt{rsapec} model fit to the Resolve FOV spectrum. 
    {Alt text: One data point and one line are plotted on a single graph with projected radius on the abscissa and integrated electron density on the ordinate. The integrated result is shown over a radius of two hundred kiloparsecs.}
    }
    \label{fig:rsapec}
\end{figure}

In our spectral analysis of Abell~2199 within the Resolve FOV,
we found that the Fe He-$\alpha$ resonance ($w$) line is weaker than predicted by the best-fit thermal model.
The best-fit $w$-to-$z$ ratio derived from the 2T (ex-wz) model is lower than the theoretical expectations from AtomDB and SPEX by slightly more than $1 \sigma$ (Figure \ref{fig:wzratio}).
Although the discrepancy in the line ratio itself is modest, the fit with the \texttt{rsapec} model indicates that the resonance line is suppressed, corresponding to an improvement of $\Delta C=25$.
Figure~\ref{fig:rsapec} shows that the data point obtained from the \texttt{rsapec} fit to the Resolve FOV spectrum is consistent with the projected electron density profile derived by integrating the deprojected electron density profiles obtained from Chandra (\citealt{johnstone2002}: 0 to $\sim$200~kpc) and XMM-Newton
(\citealt{Mirakhor2020}: $\sim$200 to $\sim$1525~kpc) observations.

\begin{figure}[htbp]
    \centering
    \includegraphics[width=0.48\textwidth]{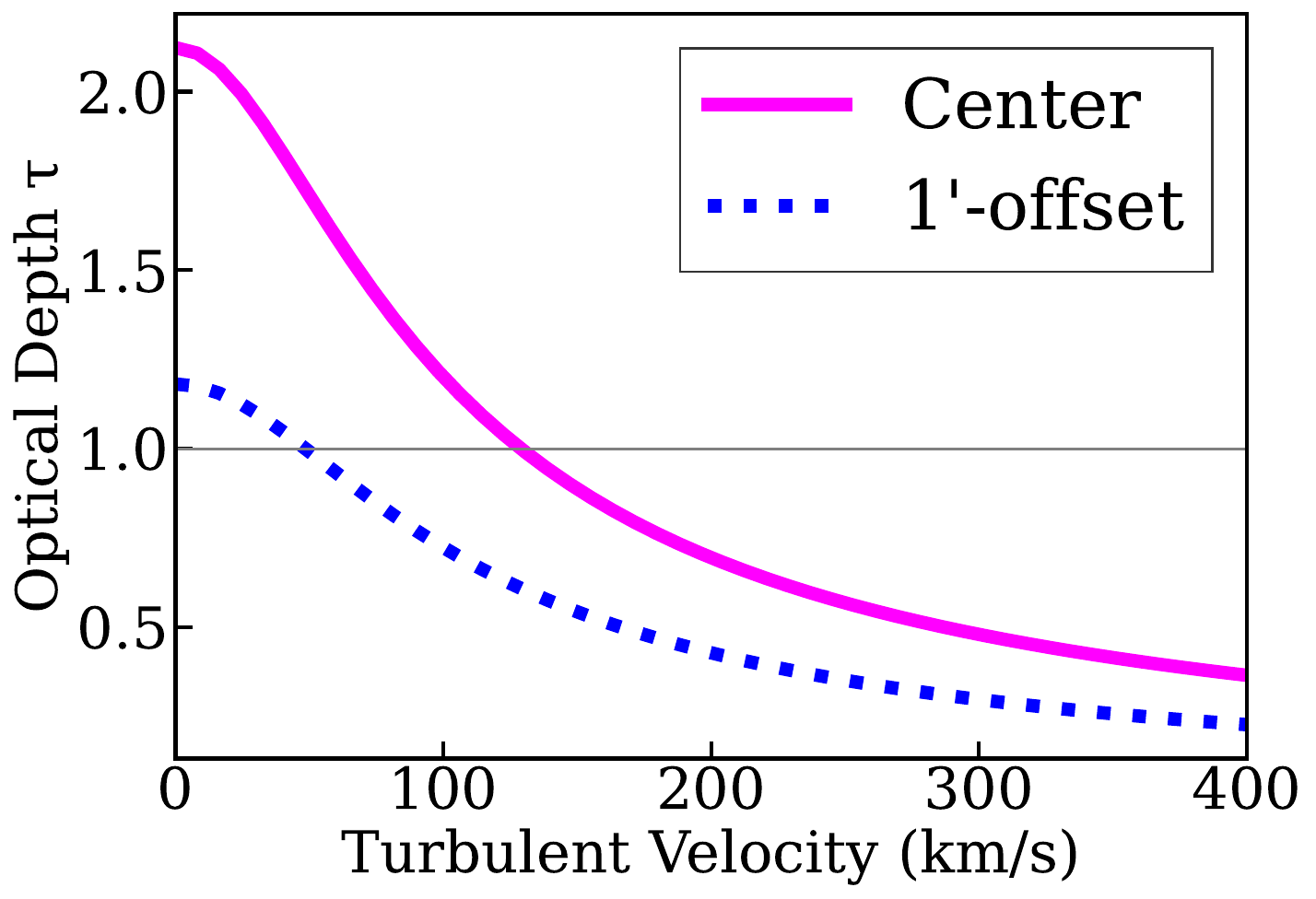}
    \caption{The optical depth at the cluster center (solid line) and 1$'$-offset (dashed line) are plotted against the turbulent velocity of the intracluster gas.
    {Alt text: Two expectation lines are plotted on a single graph with turbulent velocity on the abscissa and optical depth at the iron-helium alpha resonance line on the ordinate. The predictions are plotted for turbulence in the range from zero to four hundred kilometers per second.}
    }
    \label{fig:opticaldepth}
\end{figure}

We therefore estimated the optical depth of the Fe~He$\alpha$ resonance line using the electron density profile adopted in Figure\color{black}~\ref{fig:rsapec}, along with the Suzaku-derived temperature and abundance profiles from the same source, while using \citet{kawaharada2010} to supplement the remaining abundance data.
The results are shown in Figure \ref{fig:opticaldepth}.
At the center of A2199, the optical depth is approximately 2 in the absence of turbulent broadening. 
Including a turbulent velocity of $\sim 100~\mathrm{km~s^{-1}}$ reduces the optical depth to $\tau \sim 1.2$.
 This value is comparable to that for the center of the Perseus cluster, where a turbulent velocity of $\sim 150 ~\mathrm{km ~s^{-1}}$ leads to a suppression of the resonance line of a few tens of percent, as detected by Hitomi observations \citep{hitomicollaboration2018b}.
At 1 arcmin offset from the center, the optical depth slightly exceeds unity for zero turbulent velocity and decreases $\sim 0.7$ for a turbulent velocity of $\sim 100 ~\mathrm{km ~s^{-1}}$. 
Although these estimates depend on assumptions regarding the three-dimensional gas distribution and velocity structure, the derived optical depths indicate that resonant scattering can plausibly account for a significant fraction of the observed resonance-line deficit within the Resolve FOV.

\subsection{Anomalous Line Ratios in the Fe He$\alpha$ Triplet}

Several cluster spectra obtained with XRISM exhibit deviations in the Fe~\textsc{xxv} He$\alpha$ line ratios.
An excess at the intercombination line ($y$) in the core of Abell 2029 was reported by \citet{xrism_a2029_core}.
A similar feature was found in the Ophiuchus cluster \citep{xrism_ophiuchus} and Abell~1795 \citep{sarkar2026_a1795}, although no clear morphological structures were identified at the corresponding location.
The stacked Resolve spectrum over ten galaxy clusters also shows an excess around the $y$ line \citep{xrism_stack}.

The excess near the $y$ line in the SE region of Abell~2199 therefore provides an independent example of this anomaly in the $y$ line.  
 This localized feature is difficult to attribute solely to systematic uncertainties in the atomic data, which are estimated to be below $\sim$10\% for the relative intensities $x$ and $y$ \citep{hitomicollaboration2018c}.
Moreover, if atomic uncertainties were responsible, comparable anomalies would be expected in both regions, contrary to the observations.

Chandra data reveal a discontinuity in surface brightness located $\sim$100$''$ southeast of the X-ray peak, spatially coincident with the SE region \citep{nulsen2013}.
Across this edge, the gas density decreases by a factor of $\sim$1.7, while the temperature does not show a significant change.
Although the nature of this discontinuity remains unclear, its spatial association with the enhanced $y$ emission suggests a possible physical connection.
One possibility is an unresolved blend with Fe~\textsc{xxiv} satellite lines, such as the $q$ line, which lies close to the $y$ line in energy.
 The $q$ line is primarily produced by inner-shell excitation
of Li-like Fe, and its emissivity peaks at
$kT \lesssim 4$ keV, making its contribution more
important in relatively cool gas
\citep{fukushima2026}.  
However, the observed excess appears stronger than expected from satellite-line contributions alone, particularly given the absence of a significant temperature gradient.
Therefore, additional physical mechanisms may need to be considered.

\section{Conclusions}
In this paper, we have presented an analysis of a deep 251 ks XRISM/Resolve observation of the cool core of the galaxy cluster Abell~2199.

Our main findings are as follows:
\begin{itemize}
\item 
We measure a very small LOS velocity difference of $\sim30~\mathrm{km~s^{-1}}$ between the ICM and the BCG, and low velocity dispersion ($\sim100~\mathrm{km ~s^{-1}}$). 
The ICM in the inner $\sim100$ kpc of Abell~2199 is  therefore dynamically quiescent in the LOS direction and forms a kinematically coherent system with the BCG.
This coherent core system is offset
 from the mean cluster redshift by $\sim200~\mathrm{km~s^{-1}}$. 
These results provide an observational constraint for future simulations of the velocity field.

\item For a velocity dispersion of $\sim100~\mathrm{km~s^{-1}}$,
the estimated turbulent heating rate corresponds to $Q_{\mathrm{turb}}/Q_{\mathrm{cool}}\approx0.1 \color{black}$ for a large driving scale, as expected for sloshing, and becomes larger for smaller driving scales characteristic of AGN feedback.
Although these calculations should be regarded as order-of-magnitude estimates, they indicate that turbulent dissipation could provide a non-negligible fraction of the radiative cooling losses in the core of Abell~2199.

\item The measured velocity dispersion corresponds to
a three-dimensional Mach number of $\mathcal{M}_{\mathrm{3D}}=0.16$ and a non-thermal pressure fraction of $P_{\mathrm{NT}}/P_{\mathrm{tot}}=1.4\pm0.2\%$.
This is among the lowest in the sample of relaxed clusters observed with XRISM. Within this sample, after excluding Cygnus~A, we find no clear correlation between the velocity dispersion and AGN activity indicators such as radio luminosity or cavity power.

\item The Fe He-$\alpha$ resonance ($w$) line is weaker than predicted by the best-fit optically thin thermal model.
Since the estimated central  optical depth is approximately 2 in the absence of turbulent broadening, resonant scattering can plausibly account for the observed resonance-line deficit.

\item In the SE region, we detected a localized enhancement of the Fe~\textsc{xxv} He$\alpha$ $y$ line that cannot be solely attributed to atomic data uncertainties. This excess spatially coincides with a surface brightness discontinuity previously identified by Chandra \citep{nulsen2013}, suggesting a possible physical connection between the anomalous emission and the local gas structure. 
\end{itemize}
 
Our future spatially resolved analysis will directly compare the velocity, turbulence, and resonant scattering deficit across the core, which will provide deeper insights into the origin and nature of gas motions in this cluster.

\begin{ack}
We gratefully acknowledge the hard work of many engineers and scientists who made the XRISM mission possible. FM acknowledges financial support from the Centre national d’\'{e}tudes spatiales (CNES), France (ROR: https://ror.org/04h1h0y33) within the framework of the XRISM mission.\color{black}

\end{ack}
\bibliographystyle{mnras}
\bibliography{modified_PASJ_A2199}

\end{document}